\documentclass[amsmath,amssymb,showkeys,superscriptaddress,floatfix,aps,prd,10pt]{revtex4}
\usepackage{graphicx,epstopdf}
\usepackage[caption=false]{subfig}
\usepackage{multirow}
\usepackage{appendix}
\usepackage{xcolor}
\usepackage[colorlinks=true, urlcolor=blue, linkcolor=blue, citecolor=green]{hyperref}

\def\qslash{q\!\!\!\slash }
\def\xslash{x\!\!\!\slash }

\def\dslash{D\!\!\!\slash }
\begin{document}

\title{Nucleon's energy-momentum tensor form factors  in light-cone QCD}

\author{K. Azizi}%
\affiliation{Department of Physics, University of Tehran, North Karegar Avenue, Tehran
14395-547, Iran}
\affiliation{Department of Physics, Dogus University, Acibadem-Kadikoy, 34722 
Istanbul, Turkey}
\author{U. \"{O}zdem}%
\affiliation{Department of Physics, Dogus University, Acibadem-Kadikoy, 34722 Istanbul, Turkey}
\affiliation{Health Services Vocational School of Higher Education, Istanbul Aydin University, Sefakoy-Kucukcekmece, 34295 Istanbul, Turkey}

\begin{abstract}
We use the energy-momentum tensor (EMT) current to compute the  EMT  form factors of the nucleon  in the framework of the light cone QCD sum rule formalism. In the calculations, we employ the  most general form of the nucleon's interpolating field and use the distribution amplitudes (DAs) of the nucleon with two sets of the numerical values of the main input parameters entering the expressions of the DAs. The directly obtained results from the sum rules for the form factors are reliable at $ Q^2\geq1$~GeV$^2 $: To extrapolate the results to include the zero momentum transfer squared with the aim of  estimation of the related static physical quantities, we use some fit functions for the form factors.
 The numerical computations show that the energy-momentum tensor form factors of the nucleon can be well fitted to the multipole fit form.
We compare the results obtained for the form factors at $ Q^2=0 $ with the existing  theoretical predictions as well as  experimental data on the gravitational form factor d$_1^q(0)$. 
For the form factors M$_2^q (0)$ and J$^q(0)$ a consistency among the theoretical predictions is seen within the errors: Our results are nicely consistent with the Lattice QCD and chiral perturbation theory predictions. 
However, there are large discrepancies among the theoretical predictions on d$_1^q(0)$.  Nevertheless, our prediction is  in accord with the JLab data as well as
with the results of the Lattice QCD, chiral perturbation theory and KM15-fit. Our fit functions well define most of the JLab data in the interval $ Q^2\in[0,0.4]$~GeV$^2 $, while the Lattice results suffer  from large uncertainties in this region. As a by-product, some mechanical properties of the nucleon like the pressure and  energy density  at the center of nucleon as well as  its mechanical radius  are also calculated and their results are compared with other existing theoretical predictions.
\end{abstract}
 \date{\today}
\maketitle

\section{Motivation} 
Understanding of the inner structures of the nucleons based on the quarks and gluons degrees of freedom  is  one of the most prominent research subjects of the hadron physics. A powerful instrument to probe the nucleon's structure is to investigate
the various  form factors (FFs) of the nucleon as they carry direct information on the nucleon' substructure and geometric shape.
Indeed, the electromagnetic FFs of the nucleon unveil the way of distributions of  the charge and
 magnetization of quarks  inside the nucleon. The scalar and axial-vector
FFs encompasses information on specific viewpoints of the nucleon's substructure nature such as chiral and flavor symmetries and their breakdown.
For these reasons, extensive research has been carried out on different form factors of the nucleon for decades.
However, the gravitational  or energy-momentum tensor form factors (EMTFFs) of the nucleon have recently been received considerable attention both in theory and experiment, despite they were recommended by Pagels in 1966 \cite{Pagels:1966zza}.
These form factors cannot be extracted directly from the experiment: They can be obtained from hard exclusive reactions using the Mellin moments of certain
generalized parton distributions (GPDs). These form factors give us a tool for systematic studies of the properties of the nucleon and calculate different related observables  like energy, angular 
momentum and  pressure distributions inside the nucleon, etc.

The  matrix element of the EMT  current between the nucleon states is characterized by four form factors and 
parametrized as follows \cite{Ji:1996ek,Polyakov:2002yz}
\begin{widetext}
  \begin{eqnarray}\label{mat}
\langle N(p^\prime,s')|T_{\mu\nu}|N(p,s)\rangle&=&
 \bar{u}(p^\prime, s^\prime)\Big\lbrace M_2(Q^2)\frac{ \tilde P_\mu \tilde P_\nu}{m_N}
 +J(Q^2)\frac{i(\tilde P_\mu \sigma_{\nu\theta}+\tilde P_\nu \sigma_{\mu\theta})\Delta^\theta}{2m_N}
 +d_1(Q^2) \frac{\Delta_\mu \Delta_\nu- g_{\mu\nu} \Delta^2}{5m_N}
   \nonumber\\
  && +~ \bar c (Q^2)  m_N g_{\mu \nu} \Big\rbrace u(p,s), 
\end{eqnarray}
\end{widetext}
 where $\tilde P= (p'+p)/2$, $\Delta = p'-p$,  $\sigma_{\mu\nu}= \frac{i}{2}[\gamma_\mu, \gamma_\nu]$, $Q^2=- \Delta^2$ and  $u(p,s)$ is the spinor of the nucleon with mass $m_N$. Here, $M_2(Q^2)  $, $J(Q^2)  $, $ d_1(Q^2) $ and $\bar c (Q^2) $ are the EMTFFs of the nucleon.
The  $M_2(Q^2)$  form factor gives knowledge on the fractions of the momenta carried by the quark and gluon constituents of the nucleon.  It is also related to the  energy density distribution  inside the nucleon.
The $J(Q^2)$ form factor gives instruction about how the total angular momenta of quarks and gluons form the nucleon's spin.
The third form factor, $d_1(Q^2)$ (called the D-term), provides information on the distribution and stabilization of strong force in the nucleon.
It can be obtained by the beam charge asymmetry in deeply virtual Compton scattering (DVCS).
The negative sign of this form factor at zero-momentum transfer obtained from various theoretical studies is thought to be in connection with the spontaneous chiral symmetry  breaking \cite{Polyakov:1999gs, Kivel:2000fg, Goeke:2001tz,Jung:2014jja}.
 The quark and gluon parts of the EMT current are not conserved separately, but  their
sum is conserved.  The form factor $\bar c (Q^2)$ characterizes the order of the non-conservation of the quark part of EMT current.
This form factor is substantial to specify the distributions of the pressure forces inside the nucleon separately for quarks and gluons. This is also used  to
study the forces among quarks and gluons inside the nucleon.
Hence, the EMTFFs provide  new perspectives to the internal structure of the nucleon.
 More details can be found in a recent paper \cite{Polyakov:2018zvc}.

The EMTFFs of the nucleon have been investigated in the framework of Lattice QCD \cite{Hagler:2003jd,mathur:1999uf,  Gockeler:2003jfa, Bratt:2010jn, Hagler:2007xi,Brommel:2007sb,Negele:2004iu,Deka:2013zha}, 
chiral perturbation theory ($\chi$PT) \cite{chen:2001pva, Belitsky:2002jp, Ando:2006sk, Diehl:2006ya, Diehl:2006js, Dorati:2007bk}, instant and front form (IFF) \cite{Lorce:2018egm}, 
 Skyrme model \cite{Cebulla:2007ei,Kim:2012ts}, chiral quark soliton model ($\chi$QSM) \cite{Petrov:1998kf, Schweitzer:2002nm, Ossmann:2004bp, Wakamatsu:2005vk, Wakamatsu:2006dy, Goeke:2007fq,Goeke:2007fp, Jung:2013bya,Jung:2014jja, Jung:2015piw,Wakamatsu:2007uc}, light-cone QCD sum rules at leading order(LCSR-LO)~\cite{Anikin:2019kwi}, dispersion relation (DR)~\cite{Pasquini:2014vua} and instanton picture (IP)~\cite{Polyakov:2018exb}.
In Ref.~\cite{Hagler:2003jd}, Hagler et al. calculated the quark part of the EMTFFs of the nucleon by means of the Lattice QCD. They obtained  $ J^q \sim 0.34 \pm 0.04$ and $ M_2^q \sim 0.68 \pm 0.07$ at re-normalization scale of $\mu^2$ = 4 GeV$^2$. 
In Ref.~\cite{mathur:1999uf}, Mathur et al. calculated the quark total angular momentum of the nucleon from the quark EMTFFs on the Lattice QCD
and they found  $J^q = 0.30 \pm 0.07$ at re-normalization scale of $\mu$ = 1.74 GeV. 
In Ref.~\cite{Gockeler:2003jfa}, Gockeler et al.  performed a quenched Lattice computation of the first moment of twist-two GPDs of the proton, and assessed the total quark contribution to the spin of the proton.
They obtained  $ J^q = 0.33 \pm 0.07$, $ M_2^q = 0.55 \pm 0.11$  and $ d_1^q = -1.0 \pm 0.05$ at re-normalization scale of $\mu$ = 2 GeV. 
In Ref.~\cite{Bratt:2010jn}, Bratt et al. presented their predictions for the substructure of the nucleon from a mixed-action computation
using 2+1 flavors of asqtad sea and domain wall valence fermions.
They carried out extrapolations of their data based on various chiral effective field theory pattern at re-normalization scale of $\mu^2$ = 4 GeV$^2$.  
In Ref.~\cite{Hagler:2007xi} Hagler et al. presented a exhaustive study of the lowest moments of nucleon GPDs in 2 + 1 Lattice QCD by the help of domain wall valence quarks and refined staggered sea quarks without including the disconnected diagrams. 
They performed extrapolations of their results based on different chiral effective field theory schemes at re-normalization scale of $\mu^2$ = 4 GeV$^2$.
In Ref.~\cite{Brommel:2007sb}, Brommel et al.  reported on a calculation of nucleon's GPDs based on simulations with two dynamical non-perturbatively improved Wilson quarks with pion masses down to 350 MeV. 
They found  $ J^q = 0.226 \pm 0.013$ and $ M_2^q = 0.572 \pm 0.012$  at re-normalization scale of $\mu^2$ = 4 GeV$^2$. 
In Ref.~\cite{Deka:2013zha}, Deka et al. reported a comprehensive  computation of the quark and gluon momenta  in the
nucleon. The  computations encompass the contributions of quarks coming  from both the connected and disconnected inclusions at re-normalization scale of $\mu$ = 2 GeV.
In Ref.~\cite{Dorati:2007bk} Dorati et al. evaluated the basic properties related to the  structures of baryons  at low energies, by the help of the method of the covariant chiral perturbation theory in the baryon sector  at leading-one-loop order. They investigated the
quark-mass dependence of the isoscalar moments in the forward limit and estimated the contributions
of quarks to the total spin of the nucleon at re-normalization scale of $\mu$ = 2 GeV. 
In Ref.~\cite{Lorce:2018egm}  Lorce et al. evaluated, in details, the distributions of energy, radial pressure and tangential pressure inside the
nucleon in the framework of both the instant form and the front form of dynamics at re-normalization scale of $\mu$ = 2 GeV. 
In Ref.~\cite{Anikin:2019kwi} the author developed a method based on the light-cone sum rules at the leading order of $\alpha_s$ to compute the
gravitational form factors for the valence quark combinations in a nucleon at re-normalization scale of $\mu^2$ = 2 GeV$^2$.
In Refs.~\cite{Cebulla:2007ei,Kim:2012ts}, the EMTFFs of the nucleon are studied via the Skyrme and in-medium modified Skyrme models and they discuss how medium effects act on the form factors.
In Refs.~\cite{Petrov:1998kf, Schweitzer:2002nm, Ossmann:2004bp, Wakamatsu:2005vk, Wakamatsu:2006dy, Goeke:2007fq,Goeke:2007fp, Jung:2013bya,Jung:2014jja, Jung:2015piw,Wakamatsu:2007uc}, the EMTFFs of the nucleon are studied by means  of the chiral quark-soliton model.
It should be noted here that the Skyrme and $\chi$QSM models show the total form factors which are re-normalization scale independent.
In Ref.~\cite{Pasquini:2014vua}, Pasquini et al. presented a depiction of the D-term form factor for hard exclusive reactions, making use of unsubtracted t-channel dispersion relations at re-normalization scale of $\mu^2$ = 4 GeV$^2$.
In Ref.~\cite{Burkert:2018bqq}, Burkert et al. presented an analysis of JLab data where an experimental information on the quark
contribution to the D-term was obtained at re-normalization scale of $\mu^2$ = 1.5 GeV$^2$.
In Ref.\cite{Shanahan:2018nnv} the distributions of pressure and shear forces inside the proton are discussed via Lattice QCD computations of the EMTFFs at re-normalization scale of $\mu$ = 2 GeV.

In the present study, we compute the quark parts of the EMTFFs of the nucleon by the help of the light-cone QCD sum rules (LCSR) method,
as one of the powerful and successful nonperturbative methods in hadron physics ~\cite{Braun:1988qv, Balitsky:1989ry, Chernyak:1990ag}.
The LCSR method is based on the operator product expansion (OPE) near the light-cone and
expansion is carried out over the twists of the operators
and the features of the hadrons under study are stated with respect to the features
of the vacuum and the light-cone distribution amplitudes of the hadrons.
Since the form factors are quantities with respect to the features of the vacuum and distribution amplitudes of the hadrons,
any uncertainties in these parameters are reflected to the uncertainties of the estimations of the
form factors. This method is quite accomplished in determining the baryonic form factors at 
high $Q^2$ (see e.g.~\cite{Aliev:2004ju, Aliev:2007qu, Wang:2006su, Braun:2006hz, Erkol:2011iw, Erkol:2011qh, 
Aliev:2011ku, kucukarslan:2016xhx, Kucukarslan:2015urd, Kucukarslan:2014bla, Kucukarslan:2014mfa}).

This manuscript is organized as follows. In section \ref{secII}, we formulate and derive the
light-cone QCD sum rules for the nucleon EMTFFs.
In section \ref{secIII}, we present our numerical results for the nucleon EMTFFs. 
In section \ref{secIV}, we discuss the mechanical structure of the nucleon using
the EMTFFs.
Section \ref{secV} is reserved for the conclusions on the obtained results.
The Appendices contain: a digression on alternative notations of the EMTFFs and technical details on the model expressions.
A remark on different definitions and notations for the EMTFFs is given in Appendix A. 
The explicit expressions of the EMTFFs are moved to  Appendix B.

 \section{Formalism}\label{secII}

In order to calculate the  EMTFFs of the nucleon within LCSR,
we begin our calculations with the subsequent correlation function:

\begin{equation}\label{corf}
	\Pi_{\mu\nu}(p,q)=i\int d^4 x e^{iqx} \langle 0 |T[J_{N}(0)T_{\mu \nu}(x)]|N(p)\rangle,
\end{equation}
where $q = p'-p$ and  $T_{\mu\nu}$ is the energy-momentum tensor current.  The quark and gluon parts of the EMT current are defined as 
\begin{widetext}
\begin{eqnarray}
\label{intpol}
 T_{\mu\nu}^q (x) &=&\frac{i}{2}\bigg[\bar{u}(x)\overleftrightarrow{D}_\mu \gamma_\nu u(x) + \bar{u}(x)\overleftrightarrow{D}_\nu \gamma_\mu u (x)
  + \bar{d}(x)\overleftrightarrow{D}_\mu \gamma_\nu d(x) +\bar{d}(x)\overleftrightarrow{D}_\nu \gamma_\mu d(x) \bigg]\nonumber\\
 &&- i g_{\mu\nu} \Big[\bar{u}(x)  \big(\overleftrightarrow{\dslash}-m_u\big)u(x)
 +\bar{d}(x)\Big( \overleftrightarrow{\dslash}-m_d\Big)d(x)  \Big],\\
%
  T_{\mu\nu}^g (x) &=&\frac{1}{4}g_{\mu\nu}F^{\alpha\beta}(x)F_{\alpha\beta}(x) - F^{\mu\alpha}(x) F^{\nu}_{\alpha}(x).
\end{eqnarray}
\end{widetext}

The second part of  Eq. (\ref{intpol}) can be rewritten as~\cite{Polyakov:2018zvc}
\begin{align}
  g_{\mu\nu} \Big[\bar{u}(x)  \big(\overleftrightarrow{\dslash}-m_u\big)u(x)+\bar{d}(x)\Big( \overleftrightarrow{\dslash}-m_d\Big)d(x) \Big] &\simeq  
    g_{\mu\nu} (1+\gamma_m)\Big(m_u \bar uu + m_d \bar dd \Big),
\end{align}
where $ \gamma_m $ is the anomalous dimension of the mass operator. It should be noted here that we work in the chiral limit ($m_u = m_d = 0$).  
We also ignore the gluon fields contributions, i.e the gluonic part of the energy momentum tensor since taking into account these contributions requires knowledge of quark-gluon mixed  distribution amplitudes of the nucleon which unfortunately are not available. 
Hence, in the present study, we will deal only with  the first part of the quark part of the EMT current in  Eq. (\ref{intpol}).
The covariant derivative $ \overleftrightarrow{D}_\mu $ is defined as
$\overleftrightarrow{D}_\mu =\frac{1}{2} [ \overrightarrow{D}_\mu - \overleftarrow{D}_\mu$]
with $\overrightarrow{D}_\mu = \overrightarrow{\partial}_\mu+igA_\mu$, $\overleftarrow{D}_\mu = \overleftarrow{\partial}_\mu -igA_\mu$;
 and  $A_\mu$ is the gluon field. In the correlation function above,  $J_{N}(0)$ is the nucleon's interpolating current.
In this study, we decide on  the most general form of the interpolating current for nucleon, which is written as
 \begin{eqnarray}\label{cur}
  J_N(x) &=&2\epsilon^{abc}\Bigg[\big[u^{aT}(x) C  d^b(x)\big]\gamma_5 u^c(x)+ t\,\big[u^{aT}(x) C \gamma_5  d^b(x)\big] u^c(x)\Bigg],
\end{eqnarray}
where
$a$, $b$, $c$ are the color indices, t is an arbitrary mixing parameter,
 and $C$ is the  charge conjugation operator.
Choosing $t=-1$ reduces the above current to the  famous  Ioffe current.

In order to calculate sum rules for EMTFFs, we need to evaluate the correlator in two different languages.
First, it is computed with respect to the QCD  degrees of freedom: In terms of the parameters of the quarks and gluons and their non-perturbative interactions with the QCD vacuum.This representation is called the QCD or theoretical representation of the correlation function and it is obtained by the help of OPE in deep Euclidean space. 
In the second representation, the correlation function is calculated in terms of the hadronic  parameters like the mass, residue, form factors and other hadronic degrees of freedom. This representation of the correlation function is called the physical or hadronic representation.
 Equating the coefficients of various Lorentz structures from two different representations of the same correlation function and
carrying out a Borel transformation with the aim  of eliminating  the contributions of
the continuum and higher states, we obtain sum rules for the EMTFFs of the nucleon. To further suppress the unwanted contributions and enhance the ground state contribution we apply the continuum subtraction procedure with accompany of the quark hadron duality assumption. 

First we focus on the calculation of the hadronic side of the correlation function. To this end, we saturate the correlation function with  a complete set of the nucleon state, the integration over four-$ x $ leads to
 \begin{align}\label{phys}
 \Pi_{\mu\nu}^{Had} (p,q) =& \displaystyle\sum_{s{'}} \frac{\langle0|J_N|{N(p',s')}\rangle\langle {N(p',s')}
 |T_{\mu \nu}^q|N(p,s)\rangle}{m^2_{N}-p'^2}+...
\end{align}
where dots represent the unwanted contributions coming from the continuum and higher states.
The above relation is further simplified by introducing the following definition:
\begin{equation}\label{rezi}
 \langle0|J_N (0)|{N(p',s')} \rangle = \lambda_{N} u (p',s')
\end{equation}
where $\lambda_{N}$ is the nucleon overlap amplitude or its residue.
%
Inserting Eqs. (\ref{mat}) and  (\ref{rezi})  into  Eq. (\ref{phys}), and performing summation over the spins of the Dirac spinors,
we obtain the hadronic side of the correlation function in terms of the  hadronic properties as well as different Lorentz structures as
\begin{widetext}
\begin{align}
\label{res1}
 \Pi_{\mu\nu}^{Had}(p,q)=&\frac{\lambda_N }{{m^2_{N}-p'^2}}\Bigg[
  M_2^q(Q^2)\Bigg\{2~ p^{\prime}_\mu p^{\prime}_\nu-p^{\prime}_\mu q_\nu-p^{\prime}_\nu q_\mu+\frac{1}{2}q_\mu q_\nu
   \Bigg\}\nonumber\\
  &+J^q(Q^2)\Bigg\{-\frac{1}{2 m_N}\Big(4~ p^{\prime}_\mu p^{\prime}_\nu \qslash- p^{\prime}_\mu q_\nu \qslash  -p^{\prime}_\nu q_\mu \qslash  \Big)
 +\frac{1}{2}\Big( p^{\prime}_\nu \qslash \gamma_\mu+p^{\prime}_\mu \qslash \gamma_\nu 
  -p_\nu \gamma_\mu \qslash -p_\mu \gamma_\nu \qslash  \Big)\nonumber\\
  &-\frac{1}{4}\Big( q_\nu \qslash \gamma_\mu+q_\mu \qslash \gamma_\nu- q_\nu \gamma_\mu \qslash -q_\mu \gamma_\nu \qslash  \Big)
  +\frac{p^{\prime}.q}{2 m_N}\Big( 2~ p^{\prime}_\mu \gamma_\nu+2~ p^{\prime}_\nu \gamma_\mu+ q_\mu \gamma_\nu+q_\nu \gamma_\mu\Big)\Bigg\}\nonumber\\
  &+\frac{2}{5}d_1^q(Q^2)q_\mu q_\nu+ 2  \bar c^q(Q^2) m_N^2  g_{\mu\nu}\Bigg].
  \end{align}
\end{widetext}
 
On QCD side, we insert the explicit forms of the interpolating current for the nucleon and the EMT current into the correlation function and perform the required contractions using the Wick theorem. The resultant expression is in terms of the light quark propagator as well as the matrix elements of the quark fields sundwiched between the vacuum and nucleon states. The latter will be defined in terms of the nucleons DAs later.  As a result we get
\begin{widetext}
\begin{eqnarray}\label{corrfunc}
	\Pi_{\mu\nu}^{QCD}(p,q)&=&-\int d^4 x e^{iqx}\Bigg[\bigg\{C_{\alpha\beta} (\gamma_5)_{\gamma\delta}( \overleftrightarrow{D}_\mu (x)\gamma_\nu)_{\omega \rho} 
	+
	t(C \gamma_5)_{\alpha\beta}\, (I)_{\gamma\delta}\,( \overleftrightarrow{D}_\mu (x)\gamma_\nu)_{\omega \rho}\bigg\}\nonumber\\
	&&
      \bigg\{\Big (\delta_\sigma^\alpha \delta_\theta^\rho \delta_\phi^\beta S(-x)_{\delta \omega}
     +\, \delta_\sigma^\delta \delta_\theta^\rho \delta_\phi^\beta S(-x)_{\alpha \omega}\Big)
     \langle 0|\epsilon^{abc} u_{\sigma}^a(0) u_{\theta}^b(x) d_{\phi}^c(0)|N(p)\rangle \nonumber\\
     &&+\delta_\sigma^\alpha \delta_\theta^\delta \delta_\phi^\rho S(-x)_{\beta \omega} 
    \,\langle 0|\epsilon^{abc} u_{\sigma}^a(0) u_{\theta}^b(0) d_{\phi}^c(x)|N(p)\rangle\bigg\}
    +\, \mu \leftrightarrow \nu \Bigg],
\end{eqnarray}
\end{widetext}
where  $ I $ is the unit matrix and $S(x)$ represents the up/down   quark propagator which is given,  in the limit $m_q = 0$, as 
\begin{align}\label{pro}
	S(x)&= \frac{i\,\xslash}{2\,\pi^2 x^4}-\frac{\langle q\bar{q}\rangle}{12}\left(1+\frac{m_0^2 x^2}{16}\right)
	-ig_s\int^1_0 d\upsilon\left[\frac{\xslash}{16\pi^2 x^4} G_{\mu\nu}\sigma^{\mu\nu}
-  \frac{i\, \upsilon\, x^\mu}{4\pi^2 x^2} G_{\mu\nu}\gamma^\nu\right],
\end{align}
where, $\langle q \bar q \rangle$ is the quark condensate and  $m_0^2$ is specified with respect to the mixed quark-gluon condensate as $ m_0^2 \equiv \langle \bar q g_s G^{\mu \nu} \sigma_{\mu \nu} q\rangle / \langle \bar q q \rangle   $. 
Since the expressions proportional to  the gluon field strength tensor ($G_{\mu\nu}$) are related to the four and five-particle distribution amplitudes, the contributions of these terms are expected to be small \cite{Diehl:1998kh} and, therefore, these contributions will be neglected in our calculations.
Furthermore, the terms proportional to $\langle q\bar{q}\rangle$ are killed and they do not contribute after applying the Borel transformations. 
Hence,  only  the first term of the propagator  survives in the calculations.

 As it is clear from Eq. (\ref{corrfunc}), to proceed in the calculations, we need to know the matrix elements of the quark operators sandwiched between the vacuum and nucleon states, i. e. 
\begin{eqnarray*}
\langle 0| \epsilon^{abc} u_{\sigma}^a(a_1 x) u_{\theta}^b(a_2 x) d_{\phi}^c(a_3 x)|N(p)\rangle
\end{eqnarray*}
where $a_1$, $a_2$ and $a_3$ are some real numbers. These matrix elements are parameterized in terms of the nucleon's distributions amplitudes of different twists in the basis of the QCD conformal partial wave expansion approach~\cite{Braun:2006hz}:
\begin{widetext}
\begin{eqnarray}
 &&\langle 0|\epsilon^{abc} u_{\sigma}^a(a_1 x) u_{\theta}^b(a_2 x) d_{\phi}^c(a_3 x)|N(p)\rangle 
\nonumber \\
&&= \frac{1}{4}\Big\{ {\cal S}_1 m_N C_{\sigma \theta} \left(\gamma_5 N\right)_\phi +{\cal S}_2 m_N^2 C_{\sigma \theta} \left(\!\not\!{x} \gamma_5 N\right)_\phi 
+ {\cal P}_1 m_N
\left(\gamma_5 C\right)_{\sigma \theta} N_\phi + {\cal P}_2 m_N^2 \left(\gamma_5 C \right)_{\sigma \theta} \left(\!\not\!{x} N\right)_\phi
\nonumber \\
& &+ \Big({\cal V}_1 +\frac{m_N^2 x^2 }{4}{\cal V}_1^M \Big)  \left(\!\not\!{P}C \right)_{\sigma \theta} \left(\gamma_5 N\right)_\phi + {\cal V}_2 m_N \left(\!\not\!{P} C \right)_{\sigma \theta}
\left(\!\not\!{x} \gamma_5 N\right)_\phi  + {\cal V}_3 m_N  \left(\gamma_\mu C \right)_{\sigma \theta}\left(\gamma^{\mu} \gamma_5 N\right)_\phi
\nonumber \\
&& + {\cal V}_4 m_N^2 \left(\!\not\!{x}C \right)_{\sigma \theta} \left(\gamma_5 N\right)_\phi+ {\cal V}_5 m_N^2 \left(\gamma_\mu C \right)_{\sigma \theta} \left(i
\sigma^{\mu\nu} x_\nu \gamma_5 N\right)_\phi + {\cal V}_6 m_N^3 \left(\!\not\!{x} C \right)_{\sigma \theta} \left(\!\not\!{x} \gamma_5 N\right)_\phi
\nonumber \\
&& + \Big({\cal A}_1 +\frac{m_N^2 x^2 }{4}{\cal A}_1^M \Big)   \left(\!\not\!{P}\gamma_5 C \right)_{\sigma \theta} N_\phi + {\cal A}_2 m_N \left(\!\not\!{P}\gamma_5 C \right)_{\sigma \theta} \left(\!\not\!{x}
N\right)_\phi  + {\cal A}_3 m_N \left(\gamma_\mu \gamma_5 C \right)_{\sigma \theta}\left( \gamma^{\mu} N\right)_\phi
\nonumber \\
&& + {\cal A}_4 m_N^2 \left(\!\not\!{x} \gamma_5 C \right)_{\sigma \theta} N_\phi + {\cal A}_5 m_N^2 \left(\gamma_\mu \gamma_5 C \right)_{\sigma \theta} \left(i
\sigma^{\mu\nu} x_\nu N\right)_\phi + {\cal A}_6 m_N^3 \left(\!\not\!{x} \gamma_5 C \right)_{\sigma \theta} \left(\!\not\!{x} N\right)_\phi
\nonumber \\
&& + \Big({\cal T}_1 +\frac{m_N^2 x^2 }{4}{\cal T}_1^M \Big)  \left(P^\nu i \sigma_{\mu\nu} C\right)_{\sigma \theta} \left(\gamma^\mu\gamma_5 N\right)_\phi + {\cal T}_2 m_N \left(x^\mu P^\nu i \sigma_{\mu\nu}
C\right)_{\sigma \theta} \left(\gamma_5 N\right)_\phi \nonumber \\ 
&&+ {\cal T}_3 M \left(\sigma_{\mu\nu} C\right)_{\sigma \theta}
\left(\sigma^{\mu\nu}\gamma_5 N \right)_\phi+{\cal T}_4 m_N \left(P^\nu \sigma_{\mu\nu} C\right)_{\sigma \theta} \left(\sigma^{\mu\varrho} x_\varrho
\gamma_5 N\right)_\phi \nonumber
\\&& + {\cal T}_5 m_N^2 \left(x^\nu i \sigma_{\mu\nu} C\right)_{\sigma \theta} \left(\gamma^\mu\gamma_5 N \right)_\phi 
+ {\cal T}_6 m_N^2 \left(x^\mu P^\nu i \sigma_{\mu\nu} C\right)_{\sigma \theta} \left(\!\not\!{x} \gamma_5
N \right)_\phi \nonumber
\\&&+ {\cal T}_7 m_N^2 \left(\sigma_{\mu\nu} C\right)_{\sigma \theta}
\left(\sigma^{\mu\nu} \!\not\!{x} \gamma_5 N \right)_\phi + {\cal T}_8 m_N^3 \left(x^\nu \sigma_{\mu\nu} C\right)_{\sigma \theta} \left(\sigma^{\mu\varrho} x_\varrho
\gamma_5 N \right)_\phi \Big\}\,,\label{da-def}
\end{eqnarray}
\end{widetext}
where $N_\phi$ is the spinor of the nucleon. 
The ``calligraphic'' functions, leaving aside the terms  proportional to $x^2$ which contain
$V_1^M$, $A_1^M$ and $T_1^M$,  can be denoted in terms of the functions of the 
specific twist as:
\begin{widetext}
		    \begin{align*}
		   \mathcal{S}_1 =& S_1,                       \hspace{3.5cm} \mathcal{S}_2=\frac{S_1-S_2}{2px} ,\nonumber\\
		   \mathcal{P}_1=&P_1,                        \hspace{3.5cm}\mathcal{P}_2=\frac{P_1-P_2}{2px}\\
           \mathcal{V}_1=&V_1,                       \hspace{3.5cm} \mathcal{V}_2=\frac{V_1-V_2-V_3}{2px}, \nonumber\\
           \mathcal{V}_3=&V_3/2,                    \hspace{3.12cm} \mathcal{V}_4=\frac{-2V_1+V_3+V_4+2V_5}{4px},\nonumber\\
           \mathcal{V}_5=&\frac{V_4-V_3}{4px} ,     \hspace{2.62cm}\mathcal{V}_6=\frac{-V_1+V_2+V_3+V_4 + V_5-V_6}{4(px)^2}\\
           \mathcal{A}_1=&A_1,                      \hspace{3.5cm} \mathcal{A}_2=\frac{-A_1+A_2-A_3}{2px},\nonumber\\
           \mathcal{A}_3=&A_3/2,                              \hspace{3.12cm}\mathcal{A}_4=\frac{-2A_1-A_3-A_4+2A_5}{4px}, \nonumber\\
           \mathcal{A}_5=&\frac{A_3-A_4}{4px},                \hspace{2.5cm}\mathcal{A}_6=\frac{A_1-A_2+A_3+A_4-A_5+A_6}{4(px)^2}\\
           \mathcal{T}_1=&T_1,                                  \hspace{3.5cm} \mathcal{T}_2=\frac{T_1+T_2-2T_3}{2px}, \nonumber\\
\mathcal{T}_3=&T_7/2,                                    \hspace{3.12cm} \mathcal{T}_4=\frac{T_1-T_2-2T_7}{2px},\nonumber\\
\mathcal{T}_5=&\frac{-T_1+T_5+2T_8}{2px},                        \hspace{1.32cm}\mathcal{T}_6=\frac{2T_2-2T_3-2T_4+2T_5+2T_7+2T_8}{4(px)^2},
\nonumber\\  \mathcal{T}_7=&\frac{T_7-T_8}{4px},                 \hspace{2.62cm} \mathcal{T}_8=\frac{-T_1+T_2 +T_5-T_6+2T_7+2T_8}{4(px)^2}
\end{align*}
\end{widetext}
where $A_i, P_i, S_i, T_i$ and $V_i$ are axialvector, pesudoscalar, scalar, tensor,  and vector distribution amplitudes, respectively.  
The expansion of the matrix element is essentially an expansion in increasing twists of the distribution amplitudes. 
The distribution amplitudes $V_1$, $A_1$ and $T_1$ have twist three, $S_1$, $P_1$, $V_2$, $A_2$, $T_2$, $V_3$, $A_3$, $T_3$ and $T_7$ 
 have twist 4, $S_2$, $P_2$, $V_4$, $A_4$, $T_4$, $V_5$, $A_5$, $T_5$ and $T_8$, are of twist 5, and $V_6$, $A_6$ and $T_6$ functions have twist 6. 
The Baryon mass correction functions  $V_1^M$, $A_1^M$ and $T_1^M$ are also of twist 5.
The distribution amplitudes $G = A_i, P_i, S_i, T_i, V_i$, which are functions of $a_i px  $, can be described as 
\begin{align}
 G(a_i px) =& \int dx_1 dx_2 dx_3 ~ \delta {(x_1 + x_2 + x_3 -1)}~
  \times exp{\bigg(-ipx \sum_i {x_i a_i}\bigg)}~ G(x_i)
\end{align}
where $x_i$ with $i = 1, 2, 3$ are equivalent to the longitudinal momentum fractions carried by the quarks inside the nucleon. The explicit representations of the 
nucleon distribution amplitudes ($A_i, P_i, S_i, T_i, V_i$) can be found in Ref.~\cite{Braun:2006hz}.

The desired LCSRs for the EMTFFs are acquired by matching the coefficients of various structures from both the hadronic and QCD sides of  the correlation function in the momentum space. We use the 
structures $p^{\prime}_\mu q_\nu$, $p^{\prime}_\mu p^{\prime}_\nu \qslash$, $q_\mu q_\nu  $ and $g_{\mu \nu}$ to find the sum rules for  the form factors $M_2^q(Q^2)$, $J^q(Q^2)$, $d_1^q(Q^2)$ and $\bar c^q(Q^2)$, respectively.
For the  EMTFFs of the nucleon we obtain:
\begin{widetext}
\begin{align}\label{eqM2}
 M_2^q (Q^2)\,\frac{\lambda_N }{{m^2_{N}-p'^2}} &= \frac{m_N}{8}\Bigg[ \int_0^{1}dx_2 \frac{x_2}{(q-p\,x_2)^2}[F_1(x_2)+x_2\,F_3(x_2)]
 +\int_0^{1}dx_3 \frac{x_3}{(q-p\,x_3)^2}[F_2(x_3)+x_3\,F_4(x_3)]\nonumber\\
 & +m_N^2\int_0^{1}dx_2\frac{x_2^2}{(q-p\,x_2)^4}\,F_5(x_2) 
 +m_N^2\int_0^{1}dx_3\frac{x_3^2}{(q-p\,x_3)^4}\,F_6(x_3)\nonumber\\
 &+\int_0^{1}d\alpha \frac{1}{(q-p\,\alpha)^2} [F_7(\alpha)+F_8(\alpha)+\alpha F_9(\alpha)+\alpha F_{10}]\nonumber\\
 &+m_N^2\int_0^{1}d\alpha \frac{\alpha^2}{(q-p\,\alpha)^4}[F_{11}(\alpha)+F_{12}(\alpha)]+\alpha F_{13}(\alpha)+\alpha F_{14}(\alpha)]\nonumber\\
 &+m_N^2\int_0^1 d\beta \frac{\beta}{(q-p \beta)^4}[F_{15}(\beta)+F_{16}(\beta)+\beta F_{17}(\beta)+ \beta F_{18}(\beta)]\Bigg],\\
 \nonumber\\
J^q (Q^2)\,\frac{\lambda_N }{{m^2_{N}-p'^2}} &=-\frac{m_N}{4} \Bigg[\int_0^1 dx_2 \frac{x_2}{(q-p x_2)^2}\,F_{19}(x_2)
+\int_0^1 dx_3 \frac{x_3}{(q-p x_3)^2}\,F_{20}(x_3)]\nonumber\\
&+m_N^2\int_0^1 dx_2 \frac{x_2}{(q-p x_2)^4}[F_{21}(x_2)+x_2 F_{23}(x_2)]
+m_N^2\int_0^1 dx_3 \frac{x_3}{(q-p x_3)^4}[F_{22}(x_3)+x_3 F_{24}(x_3)]\nonumber\\
&+m_N^2 \int_0^1 d\alpha \frac{\alpha^2}{(q-p \alpha)^4}[F_{25}(\alpha)+F_{26}(\alpha)+\alpha F_{27}(\alpha)+\alpha F_{28}(\alpha)]\nonumber\\
&+m_N^2 \int_0^1 d\beta \frac{\beta}{(q-p \beta)^4}[F_{29}(\beta)+F_{30}(\beta)+\beta F_{31}(\beta)+\beta F_{32}(\beta)]\Bigg],\\
\nonumber\\
d_1^q (Q^2)\,\frac{\lambda_N }{{m^2_{N}-p'^2}} &= \frac{5\,m_N}{8}\Bigg[\int_0^1 dx_2 \frac{1}{(q-p x_2)^2}[F_{33}(x_2)+x_2 F_{35}(x_2)
+x_2^2 F_{37}(x_2)]\nonumber\\
&+\int_0^1 dx_3 \frac{1}{(q-p x_3)^2}[F_{34}(x_3)+ x_3 F_{36}(x_3)+x_3^2 F_{38}(x_3)]\nonumber\\
&+m_N^2\int_0^1 dx_2 \frac{1}{(q-p x_2)^4}[F_{39}(x_2)+x_2 F_{41}(x_2)+x_2^2 F_{43}(x_2)]\nonumber\\
&+m_N^2\int_0^1 dx_3 \frac{1}{(q-p x_3)^4}[F_{40}(x_3)+ x_3 F_{42}(x_3)+x_3^2 F_{44}(x_3)]\nonumber\\
&+\int_0^1 d\alpha \frac{1}{(q- p \alpha)^2}[ F_{45}(\alpha) + F_{46}(\alpha) + \alpha F_{47}(\alpha) + \alpha F_{48}(\alpha)   ]\nonumber\\
&+m_N^2\int_0^1 d\alpha \frac{\alpha}{(q- p \alpha)^4}[ F_{49}(\alpha) + F_{50}(\alpha) + \alpha F_{51}(\alpha) + \alpha F_{52}(\alpha) +  \alpha^2 F_{53}(\alpha)+ \alpha^2 F_{54}(\alpha) ]\nonumber\\
&+m_N^2\int_0^1 d\beta \frac{1}{(q- p \beta)^4}[ F_{55}(\beta) + F_{56}(\beta) + \beta F_{57}(\beta) + \beta F_{58}(\beta) +  \beta^2 F_{59}(\beta)+ \beta^2 F_{60}(\beta) ]\Bigg],
\end{align}
and
\begin{align}
\label{eqcbar}
\bar c^q (Q^2)\,\frac{\lambda_N }{{m^2_{N}-p'^2}} &=\frac{m_N}{16} \Bigg[ \int_0^1 d\alpha \frac{\alpha}{(q- p \alpha)^2}[ F_{61}(\alpha) + F_{62}(\alpha)] 
+ \int_0^1 d\beta \frac{1}{(q- p \beta)^2}[ F_{63}(\beta) + F_{64}(\beta)]\Bigg].
\end{align}
\end{widetext}

The explicit forms of the various  $ F $ functions  that come into view in Eqs. (\ref{eqM2}) to (\ref{eqcbar}) are
presented in the  Appendix with respect to the distribution amplitudes of the nucleon. 
For the sake of brevity,  in the  Appendix B, only the results for the $M_2^q(Q^2)$ form factor are presented, explicitly. 

The last step is to apply the Borel transformation with respect to the variable $ p'^2 $  as well as the continuum subtraction with the aim of suppression of the contributions of the higher states and continuum.  These steps are  performed by the help of the subsequent replacement rules (see e.g. \cite{Braun:2006hz}):

\begin{align}
		\int dx \frac{\rho(x)}{(q-xp)^2}&\rightarrow -\int_{x_0}^1\frac{dx}{x}\rho(x) e^{-s(x)/M^2}, \nonumber		\\
		\int dx \frac{\rho(x)}{(q-xp)^4}&\rightarrow \frac{1}{M^2} \int_{x_0}^1\frac{dx}{x^2}\rho(x) e^{-s(x)/M^2}
		+\frac{\rho(x_0)}{Q^2+x_0^2 m_N^2} e^{-s_0/M^2},
	\label{subtract3}
\end{align}
where,
\begin{eqnarray}
s(x)=(1-x)m_N^2+\frac{1-x}{x}Q^2,
\end{eqnarray}
$M^2$ is the Borel mass squared parameter and $x_0$ is the solution of the quadratic equation for $s=s_0$:
\begin{eqnarray}
x_0&=&\Big[\sqrt{(Q^2+s_0-m_N^2)^2+4m_N^2 Q^2}-(Q^2+s_0-m_N^2)\Big]/2m_N^2,
\end{eqnarray}
with $s_0$ being the continuum threshold.

One of the main input parameters in the expressions of the sum rules for EMTFFs is the nucleon's residue, $\lambda_N$.
We use the expression of this parameter, in terms of the hadronic and QCD degrees of freedom as well as the auxiliary parameters entering the calculations, calculated via  mass two-point sum rules \cite{Aliev:2011ku}. It is given as
\begin{align}
\label{residue}
\lambda_N^2 =~& e^{m_N^2/M^2} \Bigg\{\frac{M^6}{256 \pi^4} E_2(x) (5+2 t + t^2) 
- \frac{\langle \bar{q}q \rangle^2}{6} \Big[6 (1-t^2)  -
(1-t)^2  \Big] + \frac{m_0^2}{24 M^2} \langle \bar{q}q \rangle^2 \Big[12 (1-t^2) - (1-t)^2  \Big]\Bigg\}, 
\end{align}
where $x = s_0/M^2$, and
\begin{eqnarray}
\label{nolabel}
E_n(x)=1-e^{-x}\sum_{i=0}^{n}\frac{x^i}{i!}~. \nonumber
\end{eqnarray}


\section{Numerical Results}\label{secIII}

\begin{widetext}
 
 \begin{figure}[t]
\centering
  \subfloat[]{\includegraphics[width=0.3\textwidth]{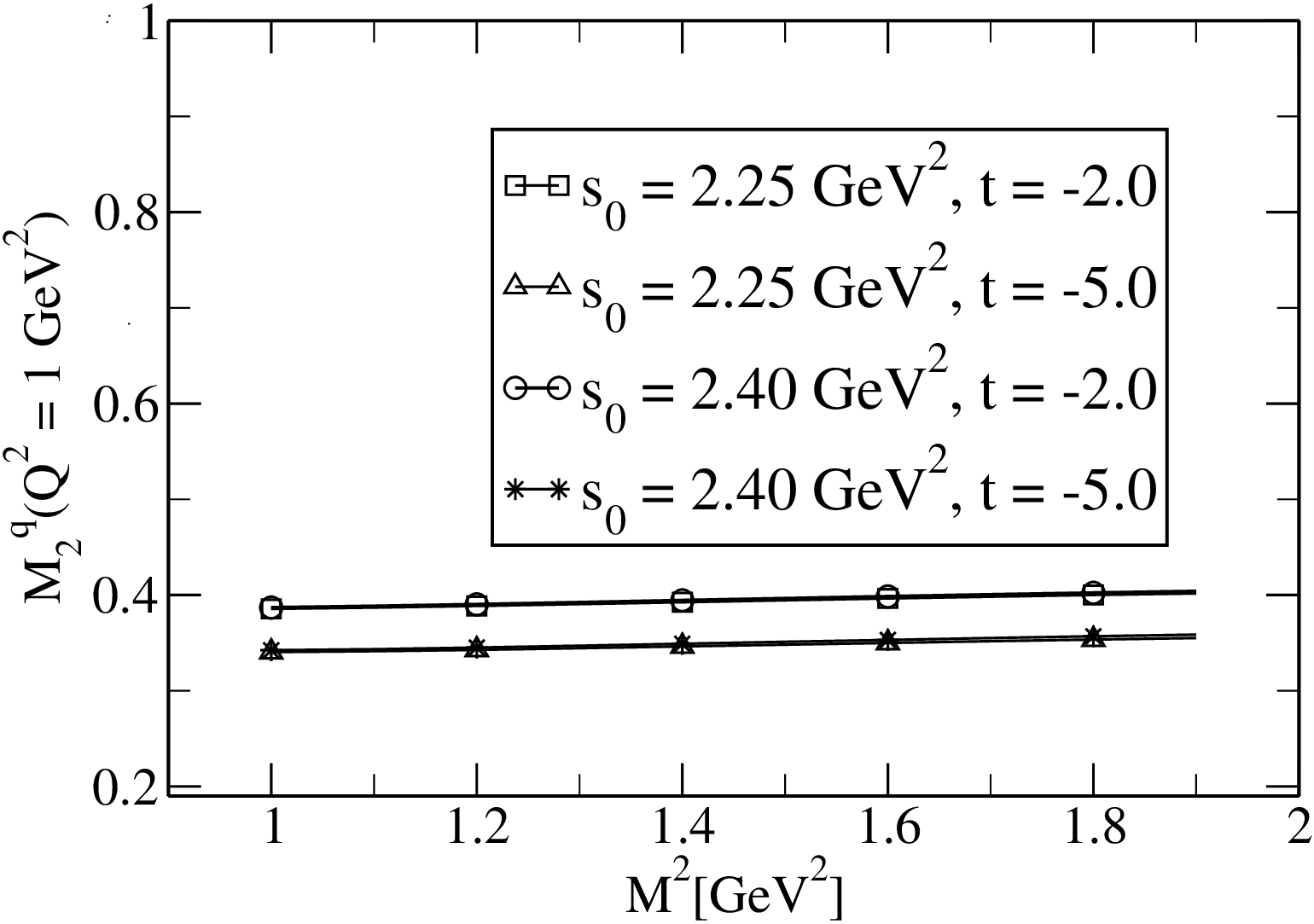}}~~~~~~~~
 \subfloat[]{ \includegraphics[width=0.3\textwidth]{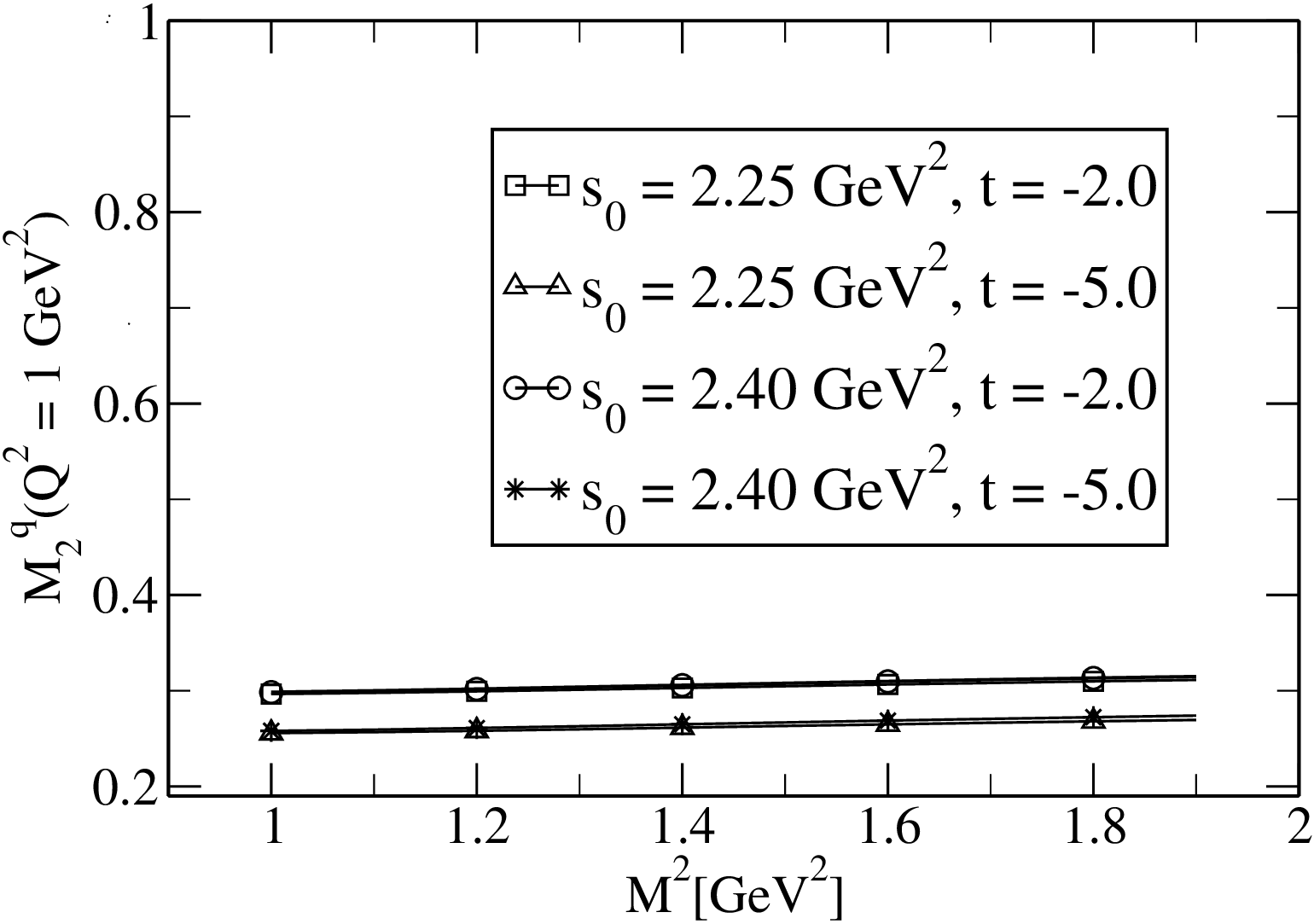}}\\
 \vspace{0.2cm}
 \subfloat[]{\includegraphics[width=0.3\textwidth]{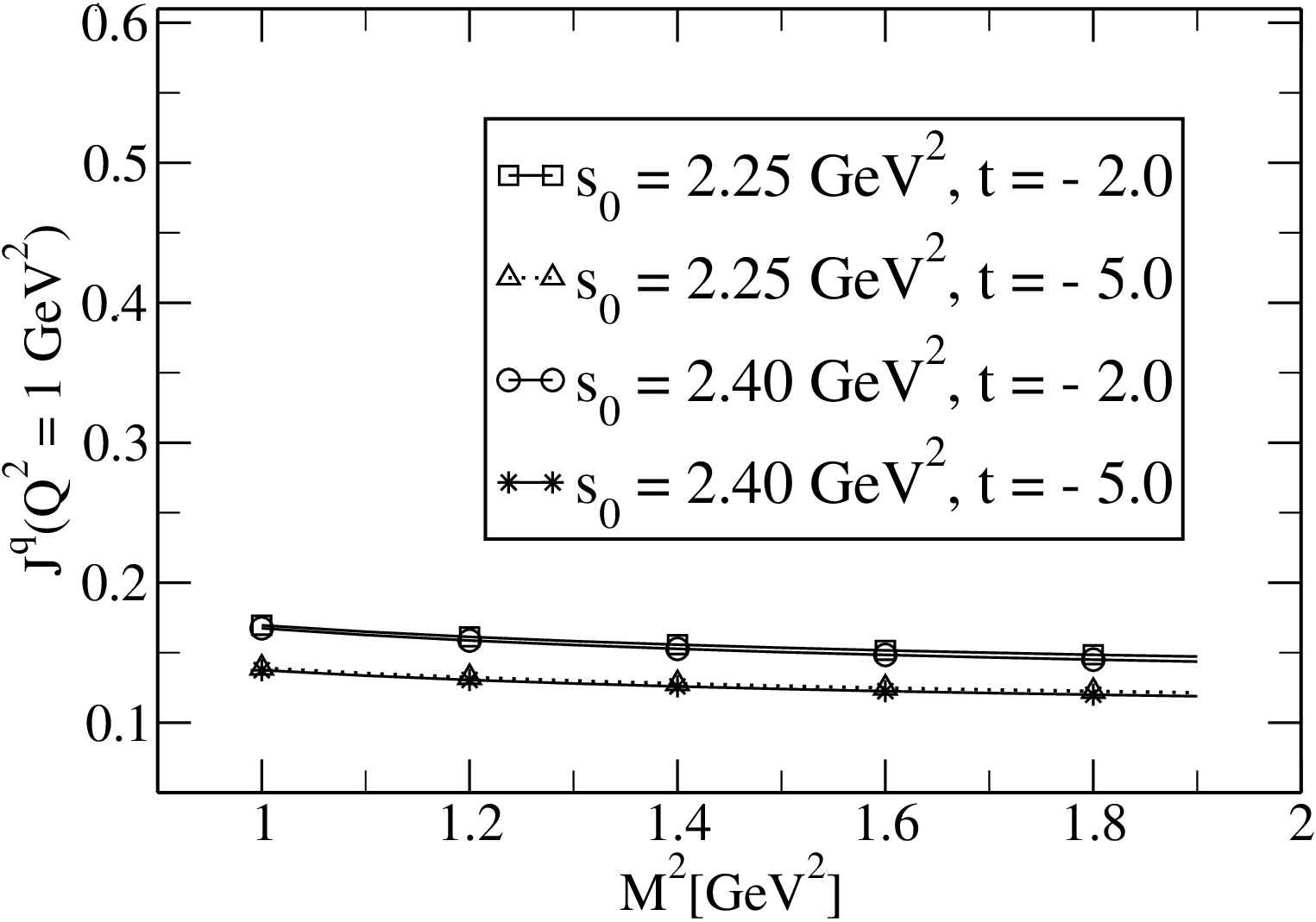}}~~~~~~~~
 \subfloat[]{ \includegraphics[width=0.3\textwidth]{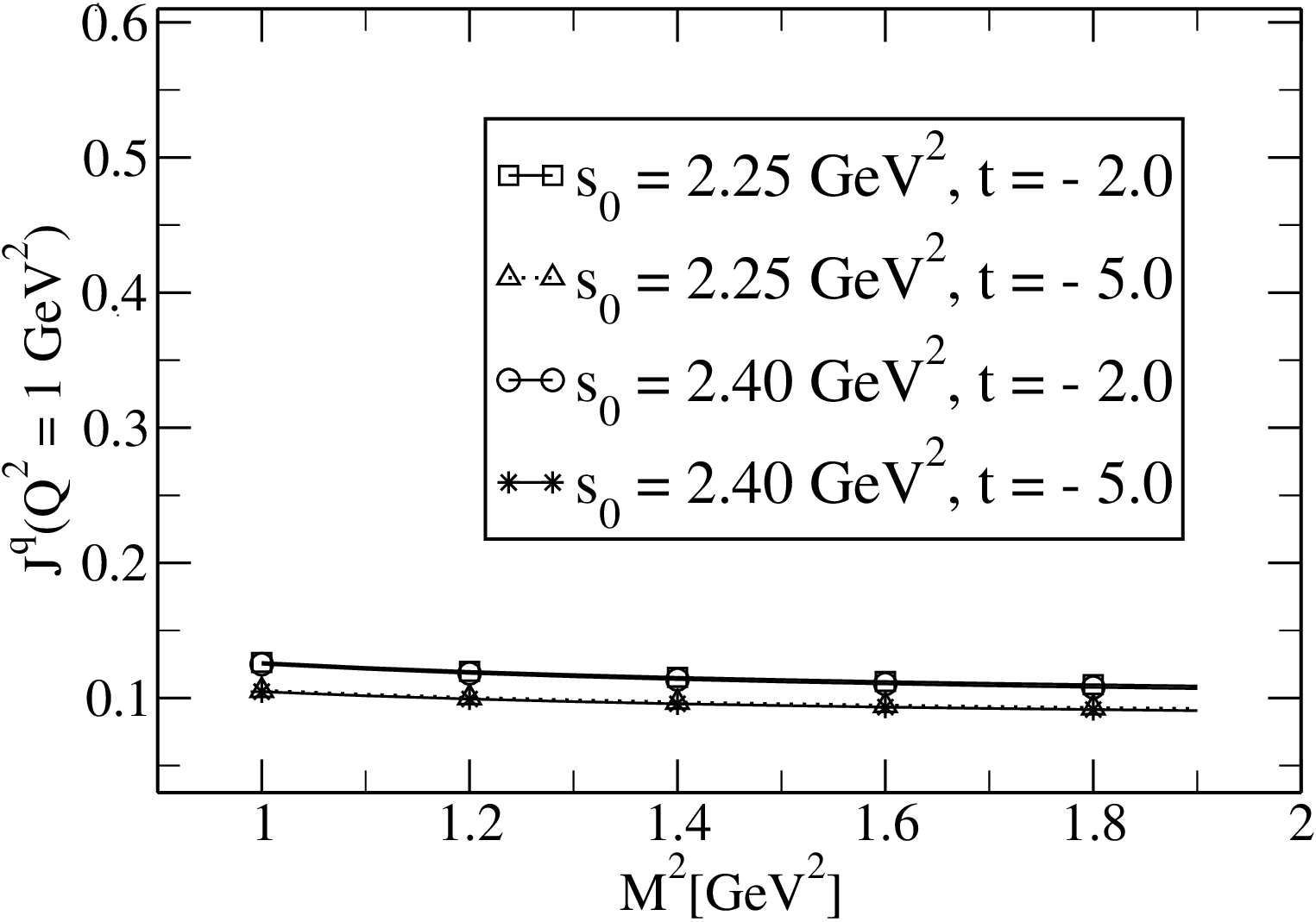}}\\
  \vspace{0.2cm}
  \subfloat[]{\includegraphics[width=0.3\textwidth]{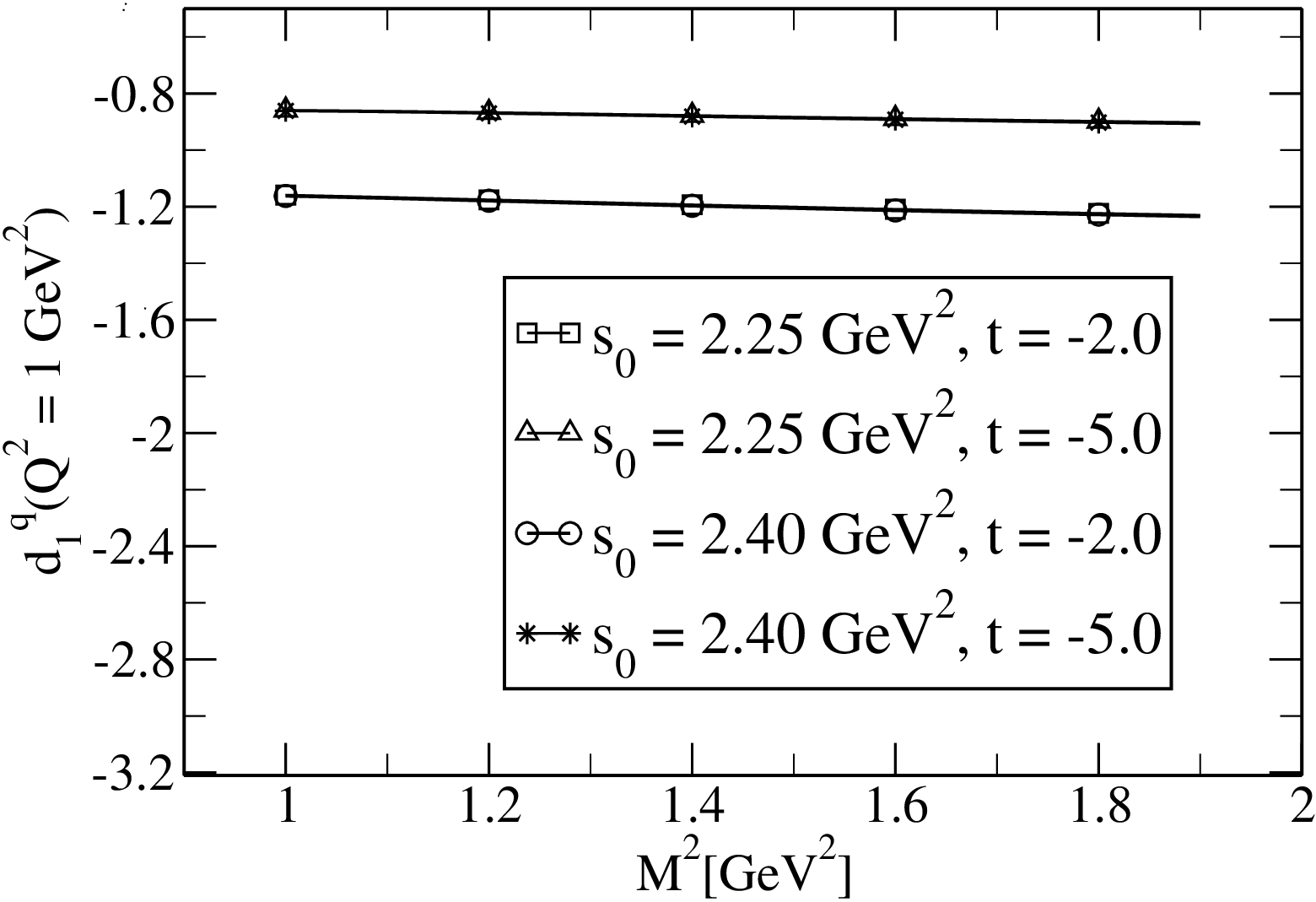}}~~~~~~~~
 \subfloat[]{ \includegraphics[width=0.3\textwidth]{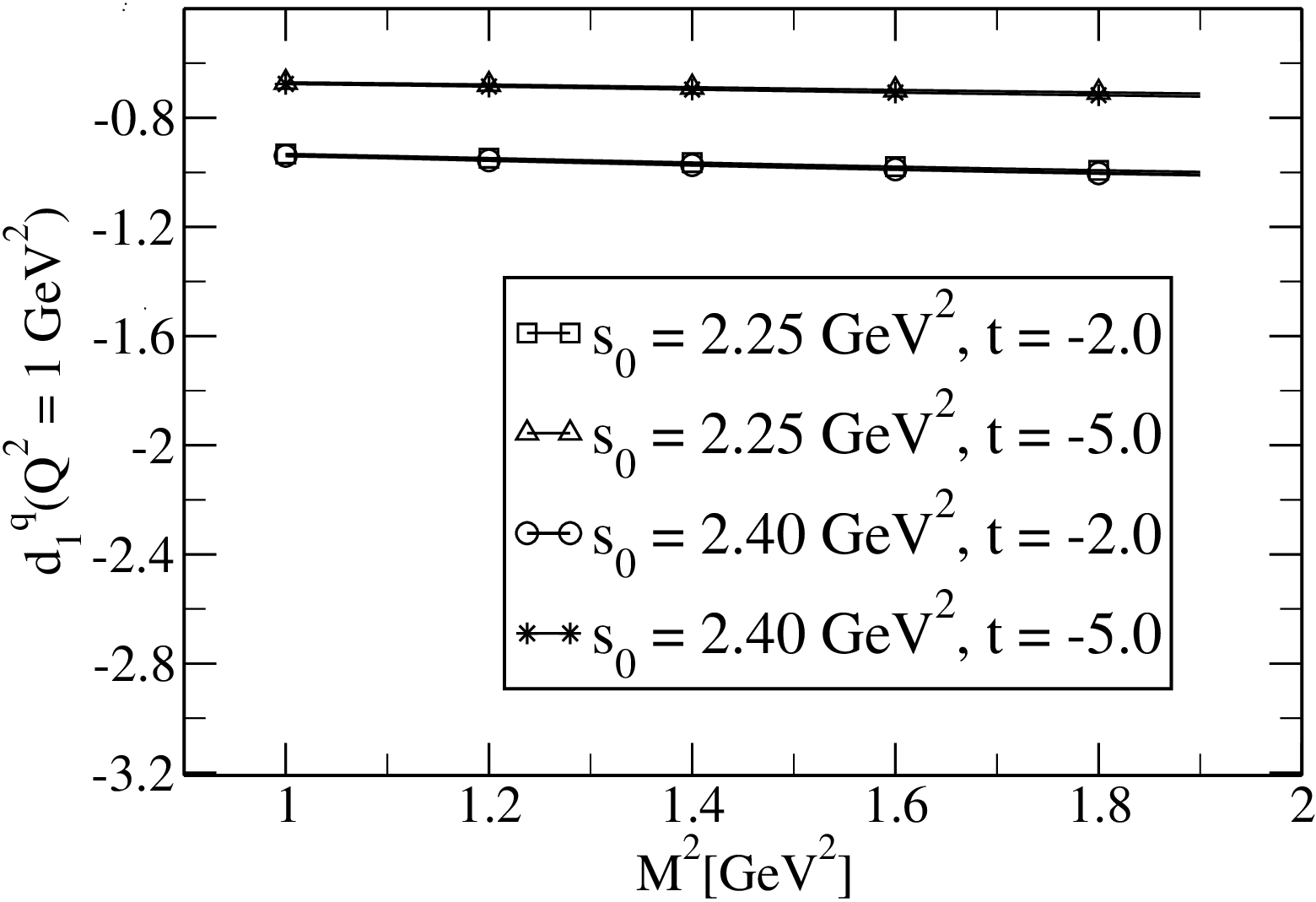}}\\
  \vspace{0.2cm}
  \subfloat[]{\includegraphics[width=0.3\textwidth]{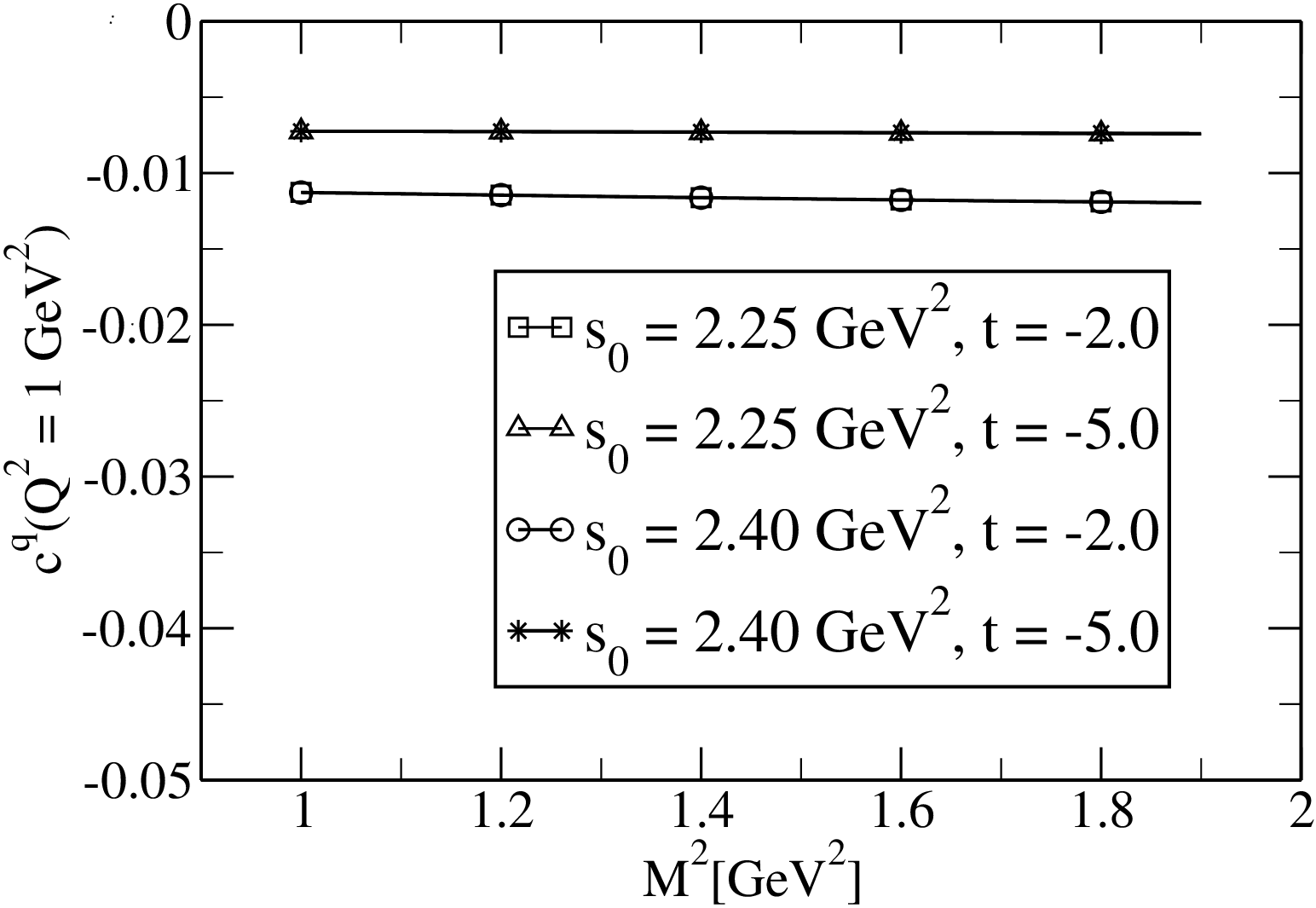}}~~~~~~~~
 \subfloat[]{ \includegraphics[width=0.3\textwidth]{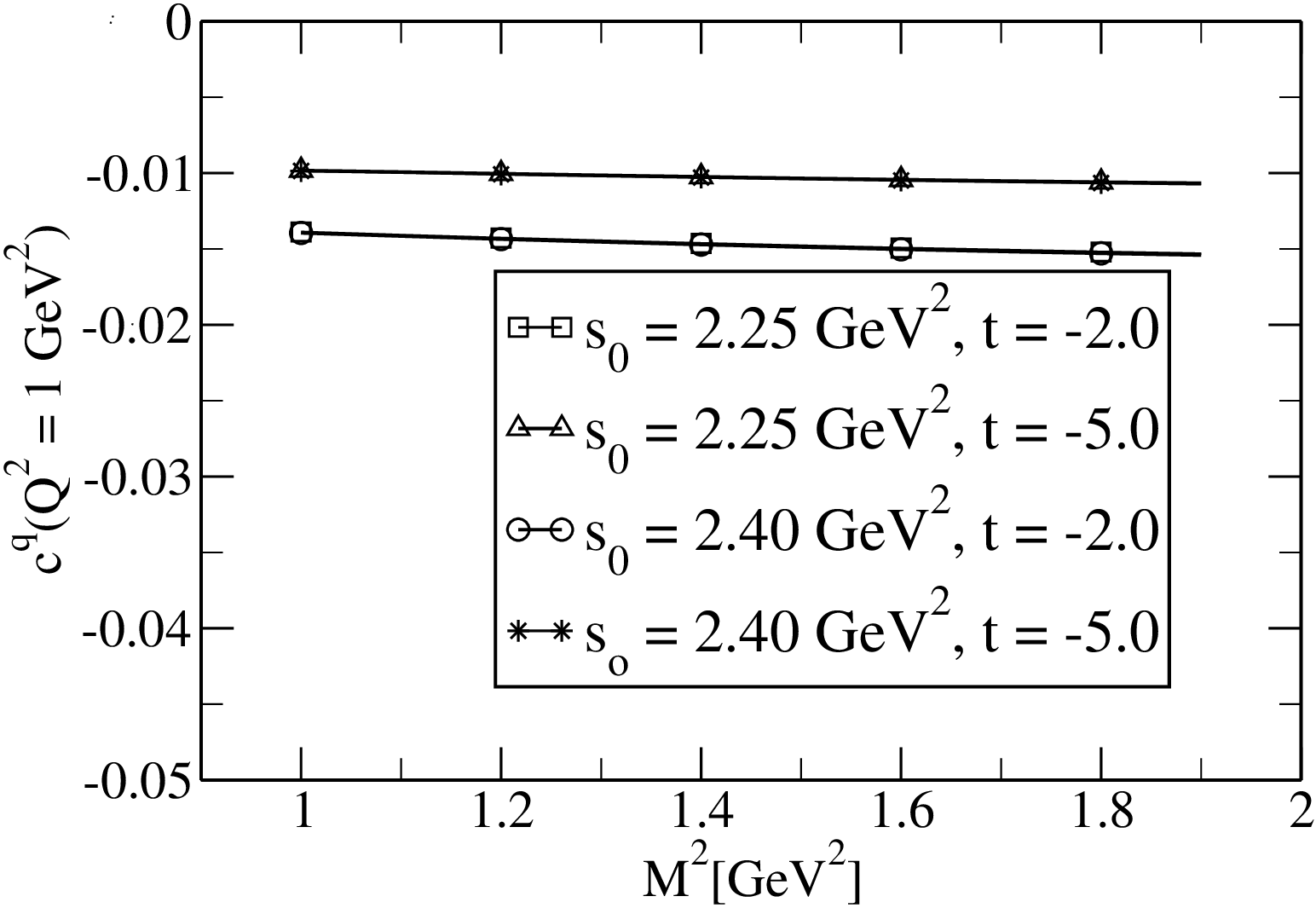}}
 \caption{The dependence of the EMTFFs of nucleon on $M^2$ at $Q^2$ = 1~GeV$^2$ and different values of $ s_0 $ and $ t $ at their working window:   (a), (c), (e) and (g) for the first set of DAs; 
 (b), (d), (f)  and (h) for the the second set of DAs.}
 \label{Msqfigs}
  \end{figure}
  
  \end{widetext}

The present section encompasses the numerical analyses of nucleon EMTFFs.
In order to obtain the numerical results of the form factors, expressions of the distribution amplitudes for nucleon are needed. 
We borrow them from Ref.~\cite{Braun:2006hz}.  
These distribution amplitudes include eight nonperturbative hadronic parameters,  which are obtained in the framework of different models. 
In further numerical computations we take into account two different sets of these parameters: 1) QCD sum rules (QCDSR) based
distribution amplitudes, where corrections to the distribution amplitudes are considered and the parameters
in distribution amplitudes are obtained from QCDSR (Set-I), 2) The
condition that the next to leading conformal spin contributions vanish, fixes
five of the eight parameters, and remaining  three parameters ($f_N$, $\lambda_1$, $\lambda_2$) are borrowed from QCDSR. 
This set is called asymptotic set (Set-II). 
The values of these parameters in two different sets are given in Table \ref{parameter_table}.
In addition we use: $m_u =m_d =0$, $m_N=0.94$~GeV,  
$\langle \bar{q}q\rangle=(-0.24\pm 0.01)^3$~GeV$^3$  and  $m_0^2=0.8 \pm 0.1$~GeV$^2$~\cite{Ioffe:2005ym}.
\begin{widetext}

\begin{table}[htp]
\renewcommand{\arraystretch}{1.3}
\addtolength{\arraycolsep}{-0.5pt}
\small
$$
\begin{array}{|l|c|c|}
\hline \hline
           &  \mbox{Set-I}                       &  \mbox{Set-II}               \\  \hline \hline
 f_N       & ( 5.0 \pm 0.5)\times 10^{-3}~GeV^2  &  ( 5.0 \pm 0.5)\times 10^{-3}~GeV^2                    \\
 \lambda_1 & (-2.7 \pm 0.9)\times 10^{-2}~GeV^2  &   (-2.7 \pm 0.9)\times 10^{-2}~GeV^2                   \\
 \lambda_2 & ( 5.4 \pm 1.9)\times 10^{-2}~GeV^2  & ( 5.4 \pm 1.9)\times 10^{-2}~GeV^2                    \\ \hline \hline
 A_1^u     & 0.38 \pm 0.15                       & 0                         \\
 V_1^d     & 0.23 \pm 0.03                       & 1/3                          \\
 f_1^d     & 0.40 \pm 0.05                       & 1/3                                            \\
 f_2^d     & 0.22 \pm 0.05                       & 4/15                                           \\
 f_1^u     & 0.07 \pm 0.05                       & 1/10                                          \\ \hline \hline
\end{array}
$$
\caption{The numerical values of the main input parameters entering the expressions of the nucleon's DAS.
	}
\renewcommand{\arraystretch}{1}
\addtolength{\arraycolsep}{-1.0pt}
\label{parameter_table}
\end{table}

\end{widetext}
There are three auxiliary parameters of the QCDSR to be fixed: the continuum threshold $s_0$, the  Borel mass parameter $M^ 2$ and the mixing parameter $ t $. 
The continuum threshold starts from the point, where the excited states and continuum contribute to the correlation function. 
We use the continuum threshold in the range $s_0 \simeq (2.25 - 2.40) $ GeV$^2$, which is pretty much fixed in the literature from the nucleon spectrum analyses. 
 The working region of $M^2$ is decided to be in the interval $1.0$ GeV$^ 2$ $\leq M^2 \leq 2.0$ GeV$^ 2 $. As it can be seen from Fig. \ref{Msqfigs} (as an example), 
the results of the form factors are roughly independent of the Borel parameter in the interval $1.0$ GeV$^ 2$ $\leq M^2 \leq 2.0$ GeV$^ 2 $. 
We include into the final results the errors coming from the variations of the physical observables with respect to the auxiliary parameters, which remain below the limits accepted by the sum rules computations. The next step is to specify the optimal mixing parameter $ t $.
%
Our numerical calculations indicate that the form factors are not sensitive to
cos$\theta$ (with $ t= $ tan$\theta$) when it varies in the region -0.2 $\leq cos\theta \leq$ -0.4.
 We see that the famous Ioffe current for the nucleon, which corresponds
to the choice cos$\theta \simeq -0.71$  remains out of the reliable the working region.

Figure \ref{Qsqfigs} shows the changes of the energy-momentum tensor form factors of the nucleon with respect to the momentum transfer squared $Q^2$.
As is also clear from this figure, the sum rules for the EMTFFs give reliable results for $Q^2\geq 1.0$ GeV$^2$. To extrapolate the results to the smaller points as well as $Q^2=0$, which enables us to compute the static properties of the nucleon, we use some fit functions such that the results of fit functions coincide with the LCSRs predictions at the region  
$Q^2\geq 1.0$ GeV$^2$.
%
%
%
 %
 Our numerical computations show that, the EMTFFs of the nucleon can be well described
by the multipole fit functions defined as \cite{Anikin:2019kwi}
%
 %
\begin{equation}
{\cal F}(Q^2)= \frac{{\cal F}(0)}{\Big(1+ m_{p}\,Q^2\Big)^p}.
\end{equation}
\begin{widetext}
 
\begin{table}[t]
\hspace*{-0.5cm}
\begin{tabular}{ |l|c|c|c|c|c|c|}
\hline
&\multicolumn{3}{|c|}{Results of set-I} &\multicolumn{3}{|c|}{Results of set-II}\\
\hline\begin{tabular}{c}Form Factors \end{tabular}& \begin{tabular}{c} ${\cal F}(0)$ \end{tabular} & \begin{tabular}{c}$m_{p}$(GeV$^{-2})$ \end{tabular}& \begin{tabular}{c}p \end{tabular}
&${\cal F}(0)$ & $m_{p}$(GeV$^{-2})$ & p \\ \hline\hline
        M$_2^q(Q^2)$       &$0.79\pm 0.10$  &$0.95 \pm 0.05 $ &$3.60 \pm 0.15$ & $0.74 \pm 0.12$ &$0.90 \pm 0.05$ &$3.40 \pm 0.10$ \\
        J$^q(Q^2)$         &$0.36 \pm 0.10$ & $0.90 \pm 0.05 $ &$3.20 \pm 0.10$ & $0.30 \pm 0.08$ & $0.83 \pm 0.05$ &$3.15 \pm 0.10$ \\
        d$_1^q(Q^2)$       &$ -2.29 \pm 0.58$& $0.95 \pm 0.05 $ & $3.45 \pm 0.15 $ & $-2.05 \pm 0.40$ &$0.90 \pm 0.05$ &$3.40 \pm 0.10$  \\
       $\bar c^q(Q^2)$    &$-(2.1\pm 0.8) \times 10^{-2}$ &$1.05 \pm 0.17$ &$3.30 \pm 0.10$&$-(2.5 \pm 0.7)\times 10^{-2}$ &$1.00 \pm 0.12$ &$3.20 \pm0.10$\\
\hline \hline
\end{tabular}
\caption{The numerical values of multipole fit parameters ${\cal F}(0)$,  $m_{p}$ and $ p $ for different EMTFFs obtained using the set-I and set-II distribution amplitudes.}
	\label{fit_table1}
\end{table}

\end{widetext}
 The  values of the fit parameters, i. e. the form factors at $ Q^2=0 $, $m_{p}$ and $ p $ for different EMTFFs obtained from the sum rules analyses are shown in Table \ref{fit_table1}.
%
The errors in the presented results are due to the variations in the computations of the working regions of $M^2$, $s_0$ and
mixing parameter $ t $ as well as the uncertainties in the values of the input parameters and the nucleon distribution amplitudes.
 Although the central values of the form factors at $ Q^2=0 $ obtained via two sets of distribution amplitudes differ slightly
from each other but they are consistent within the presented errors. 
%
%
%

The individual quark and gluon EMTFFs are not re-normalization scale independent.
The numerical values of distribution amplitudes are used at the scale $\mu^2 = 1$  GeV$^2$ in Ref. \cite{Chernyak:1984bm}. 
In the present study our estimations correspond to $\mu^2 = 1.0$  GeV$^2$, as well. Different sources use different scales to calculate the EMTFFs. In order to compare the results, we should bring them in the same re-normalization scale. To this end, we use the evolution equations, for the  form factors under consideration, calculated  in Refs. ~\cite{Hatta:2018sqd,Tanaka:2018nae}.
%
\begin{widetext}

\begin{figure}[t]
\centering
 \subfloat[]{\includegraphics[width=0.3\textwidth]{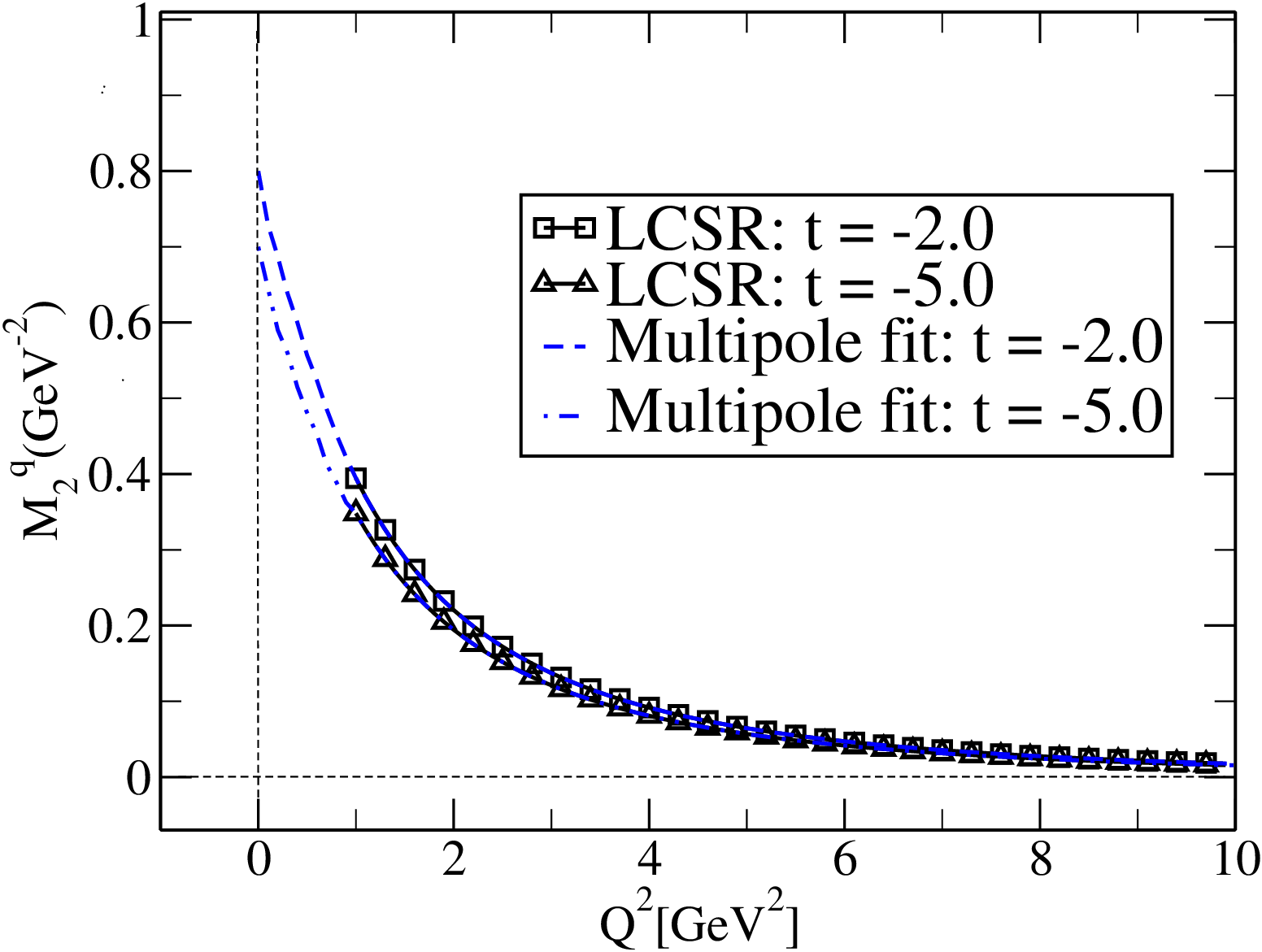}}~~~~~~~~
 \subfloat[]{ \includegraphics[width=0.3\textwidth]{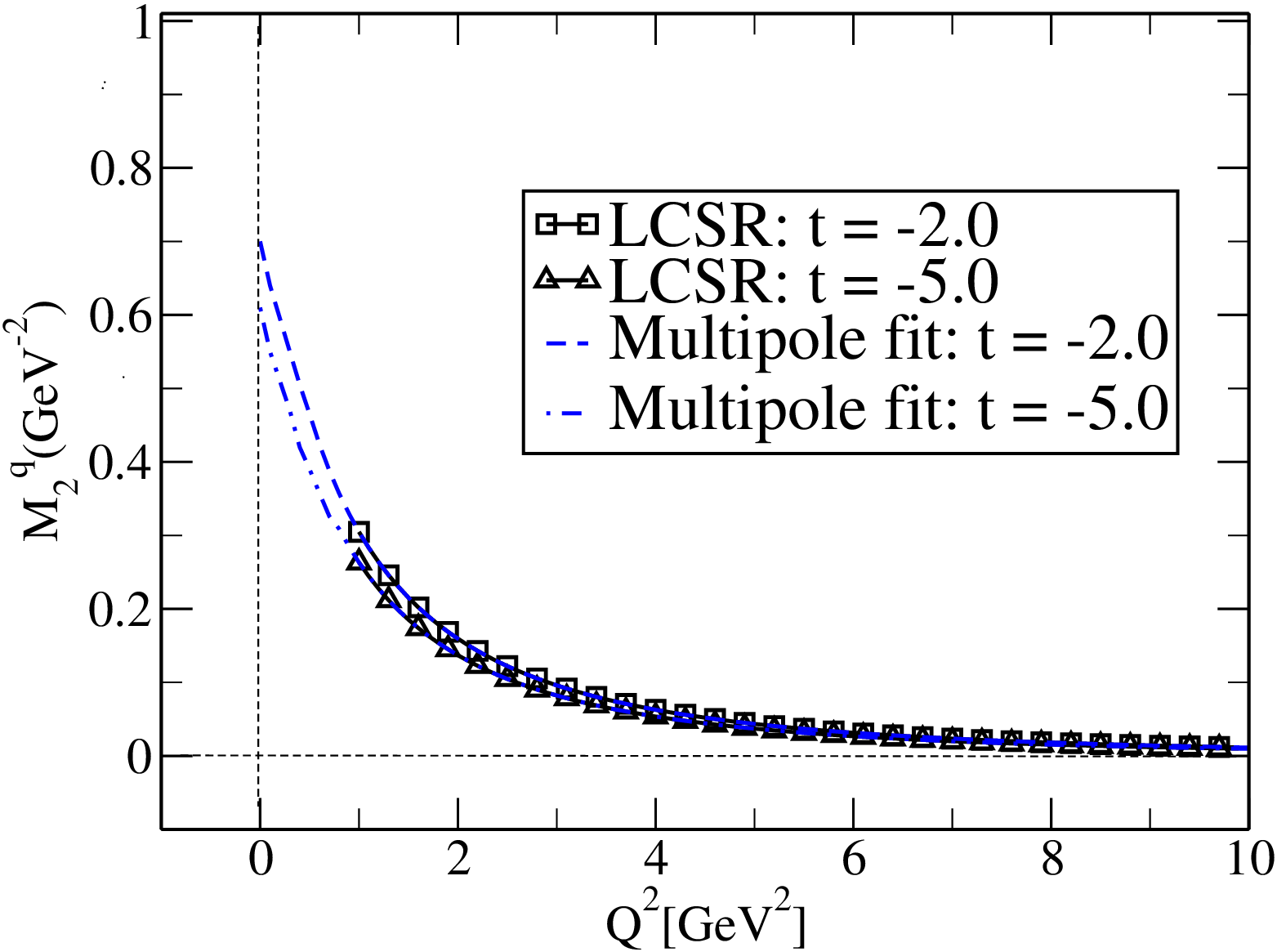}}\\
   \vspace{-0.1cm}
    \subfloat[]{\includegraphics[width=0.3\textwidth]{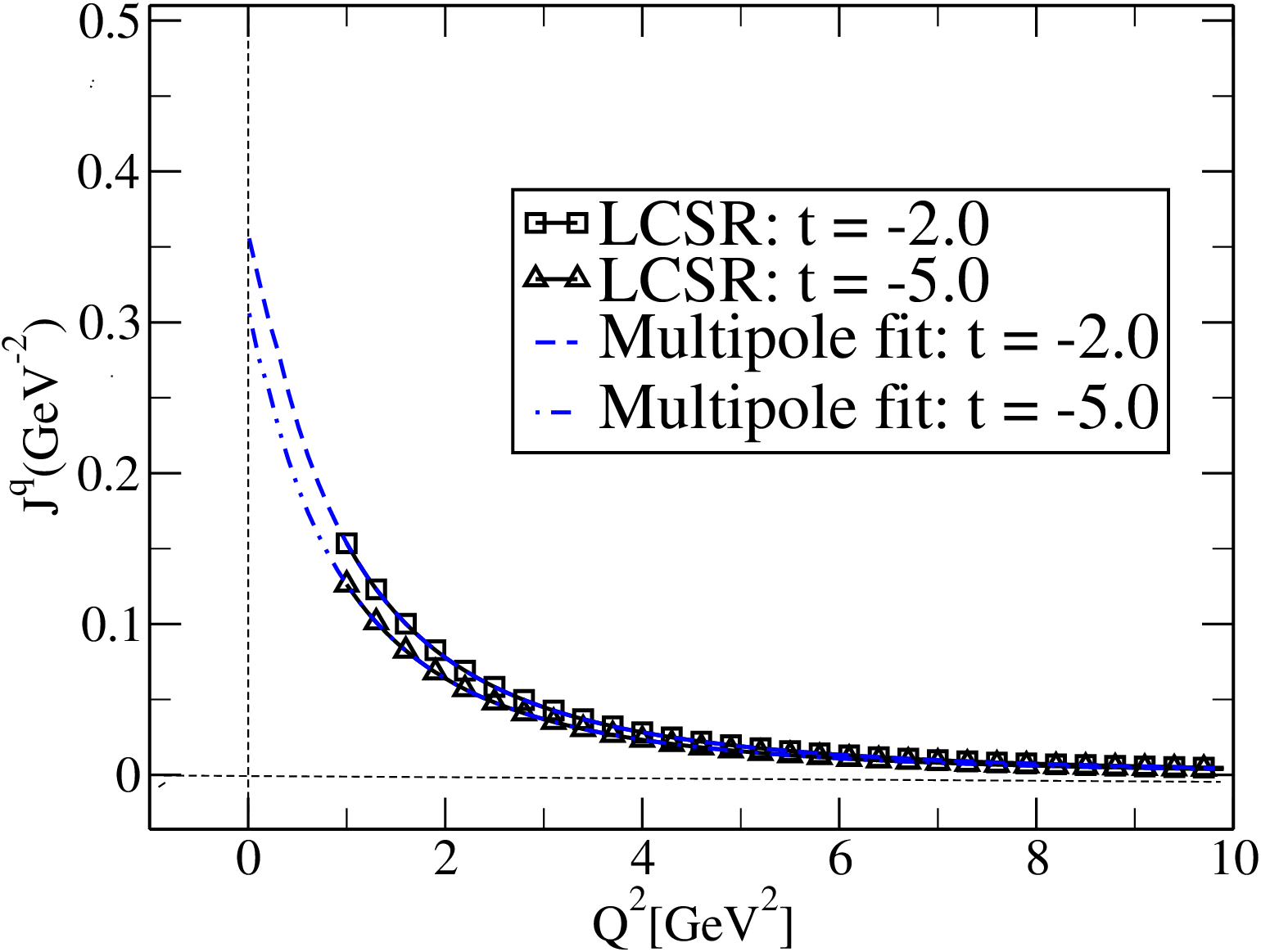}}~~~~~~~~
 \subfloat[]{ \includegraphics[width=0.3\textwidth]{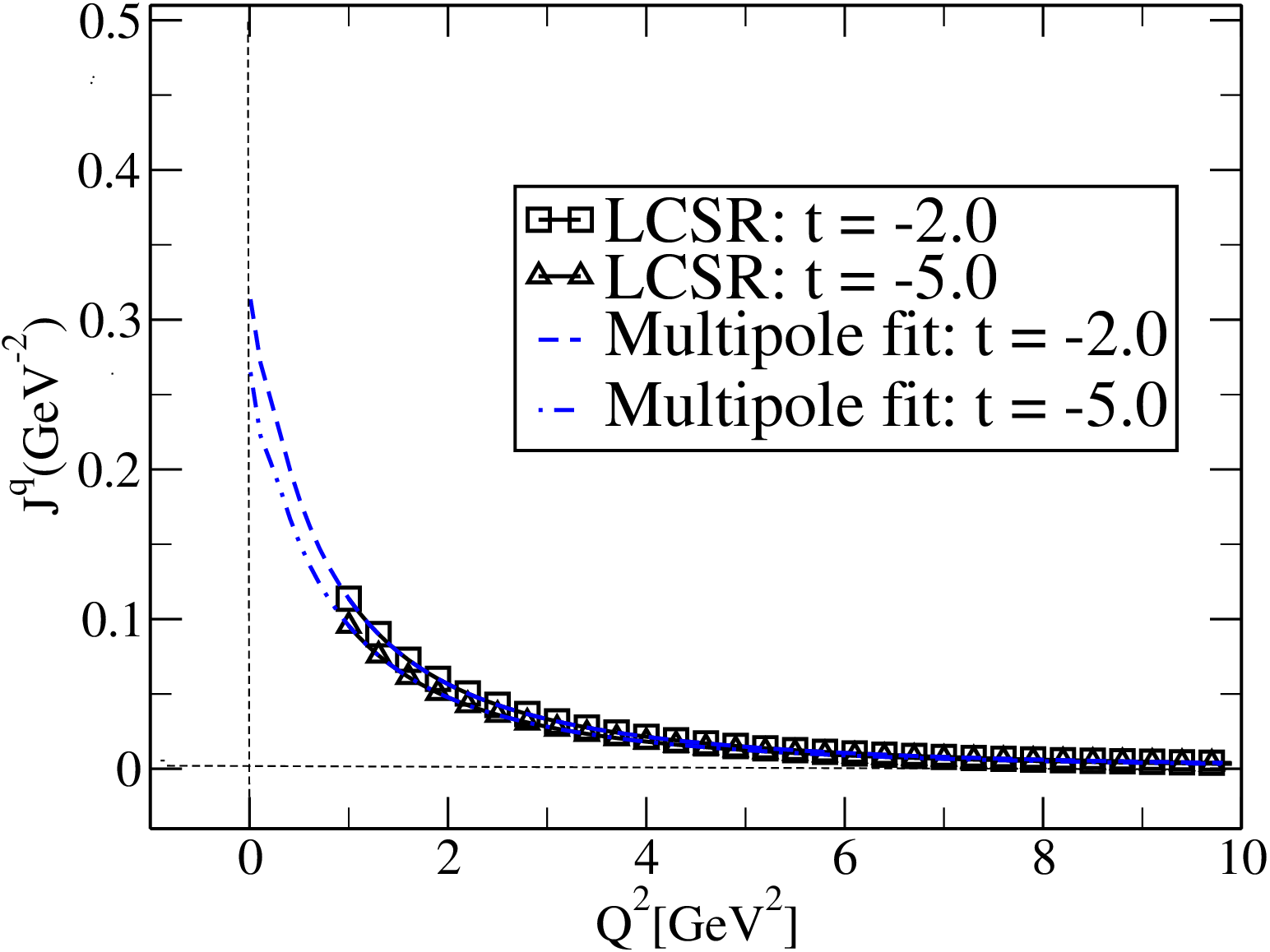}}\\
 \vspace{-0.1cm}
    \subfloat[]{\includegraphics[width=0.3\textwidth]{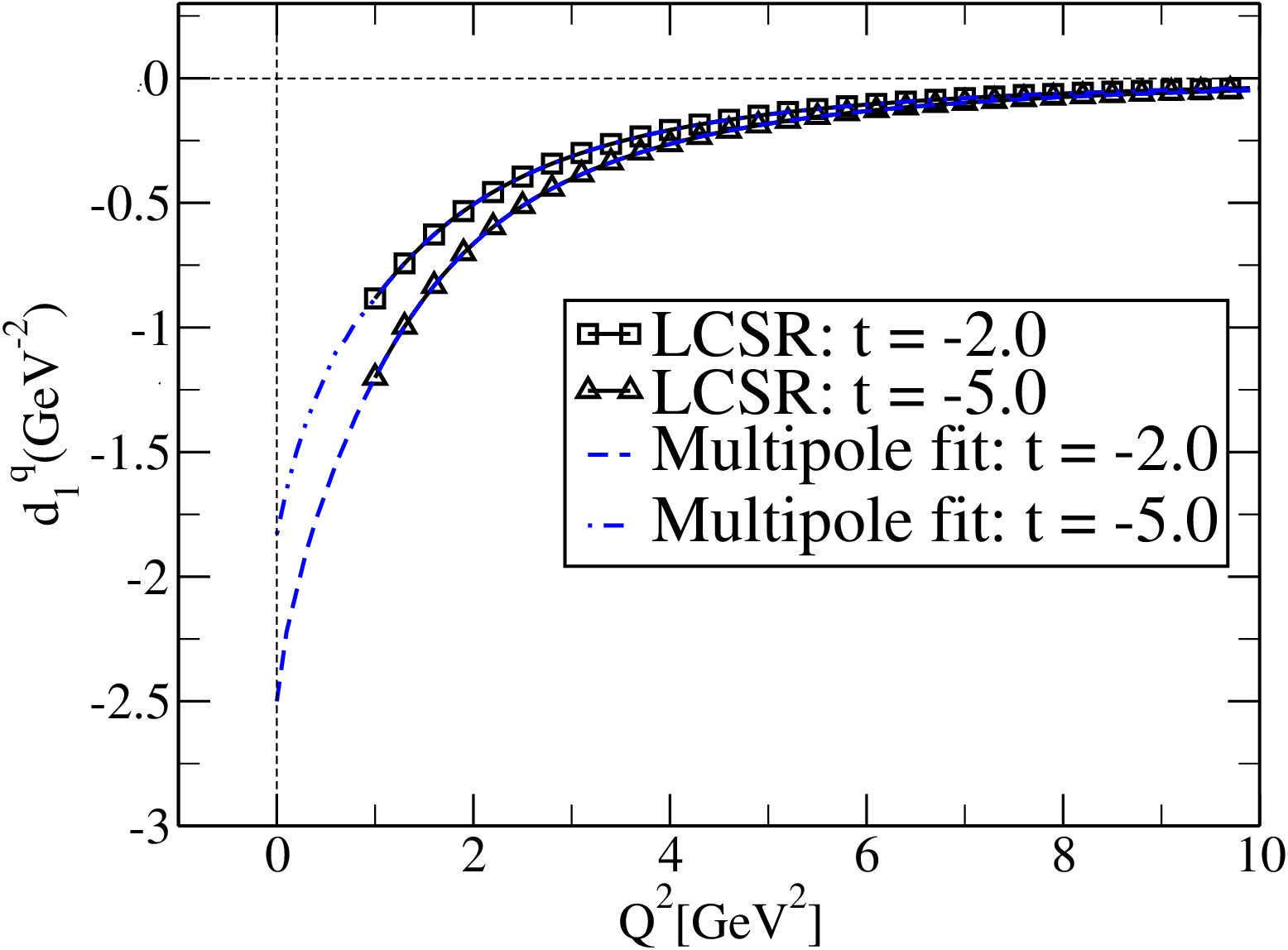}}~~~~~~~~
 \subfloat[]{ \includegraphics[width=0.3\textwidth]{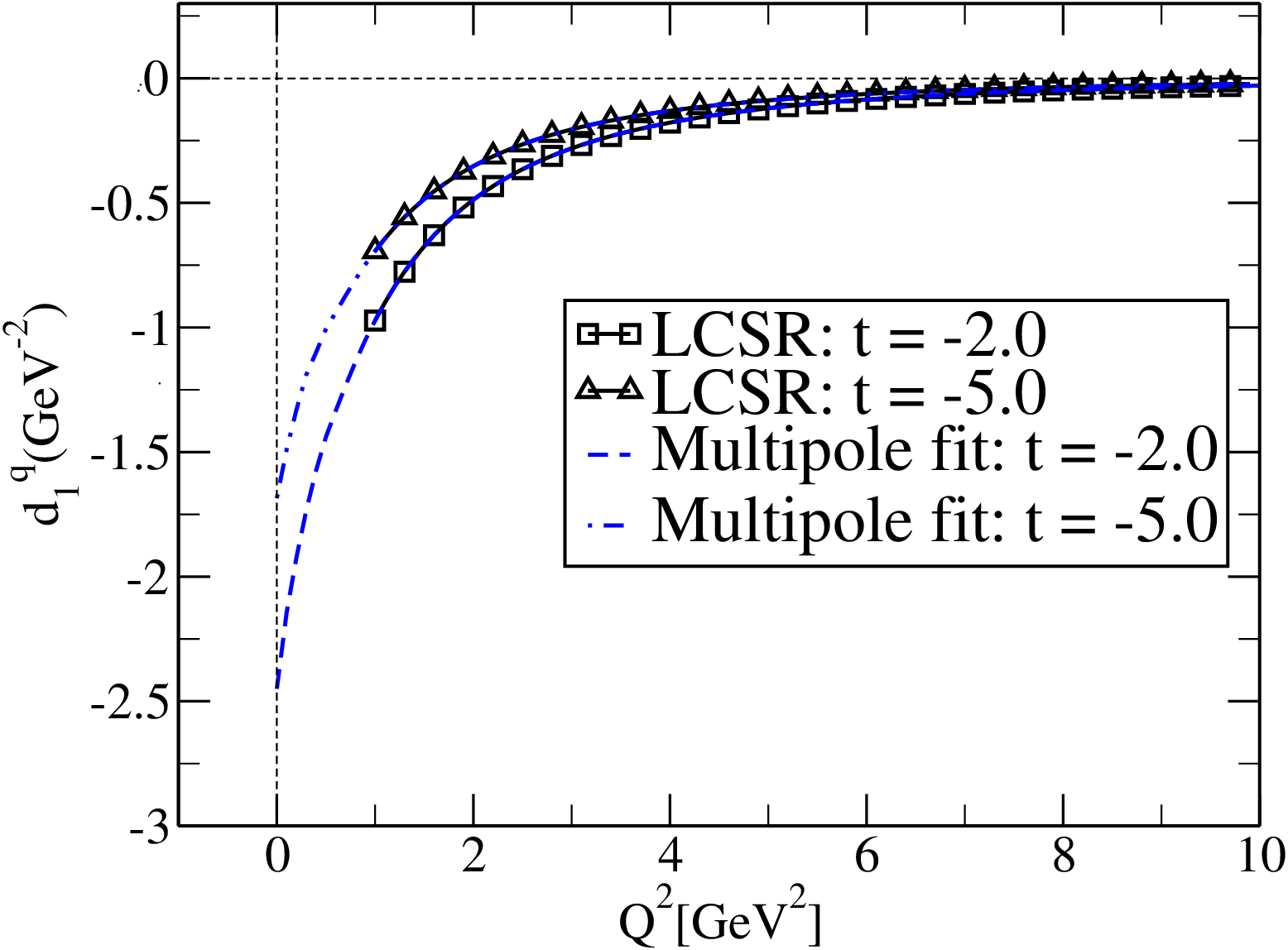}}\\
  \vspace{-0.1cm}
 \subfloat[]{\includegraphics[width=0.3\textwidth]{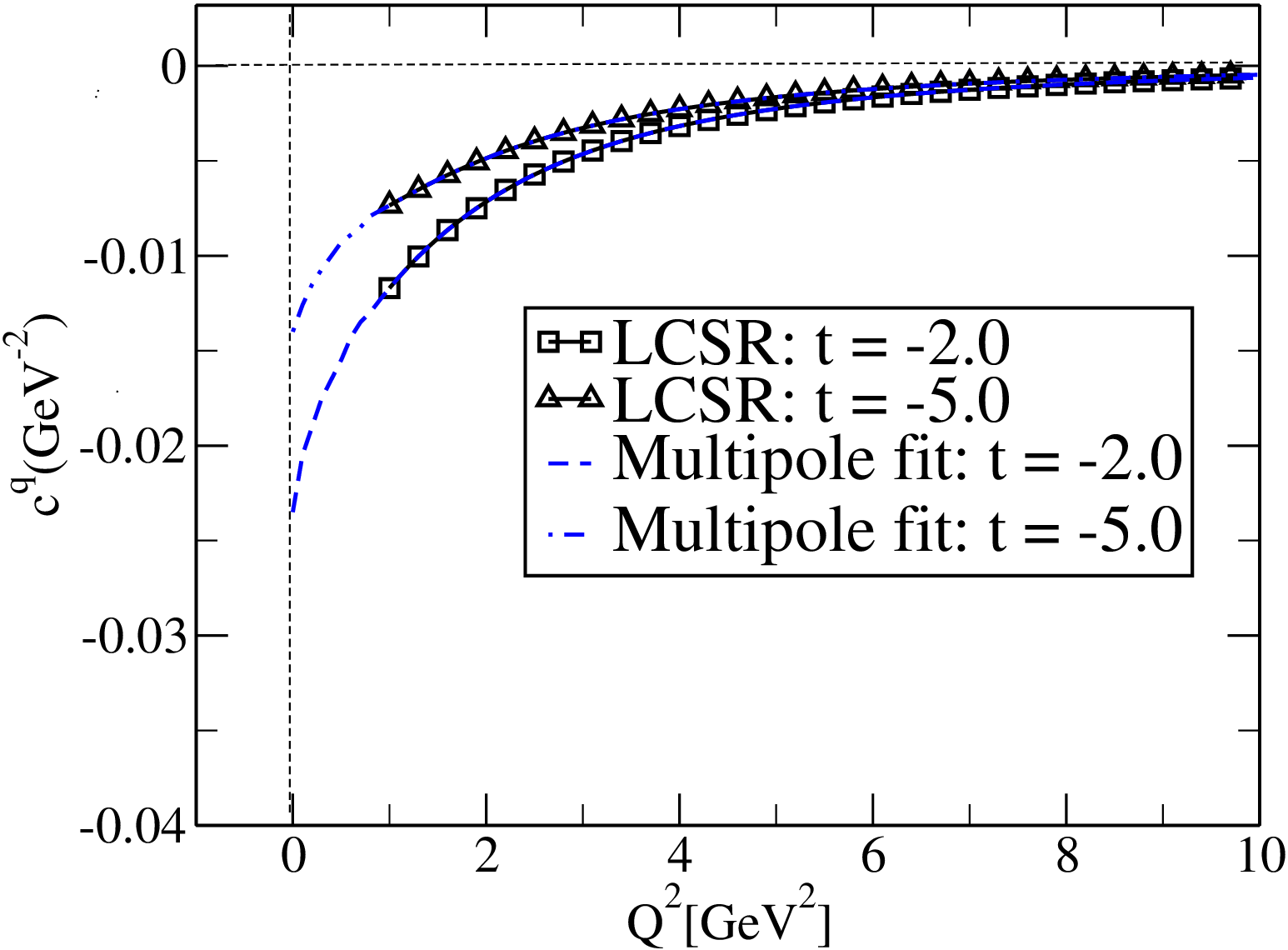}}~~~~~~~~
 \subfloat[]{ \includegraphics[width=0.3\textwidth]{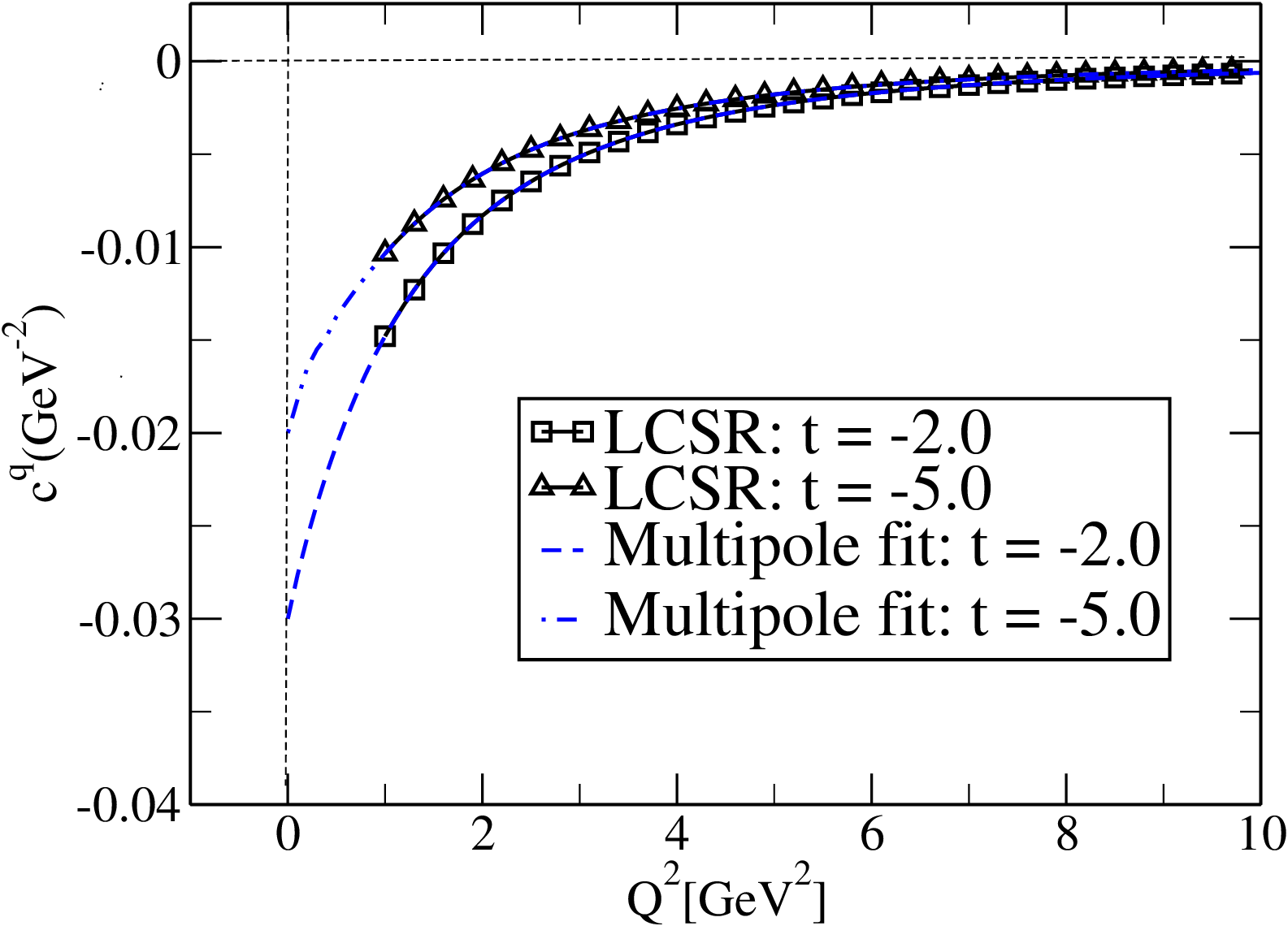}}
 \caption{The dependence of the EMTFFs of nucleon on $Q^2$ at $M^2$ = 1.5~GeV$^2$,
$s_0$ = 2.25~GeV$^2$ and  different values of $ t $:   (a), (c), (e) and (g) for the first set of DAs; 
 (b), (d), (f)  and (h) for the the second set of DAs. The dashed and dot-dashed curves show the results of the fit functions of the  multipole form.}
 \label{Qsqfigs}
  \end{figure}

 \begin{figure}[t]
\centering
 \subfloat[]{\includegraphics[width=0.3\textwidth]{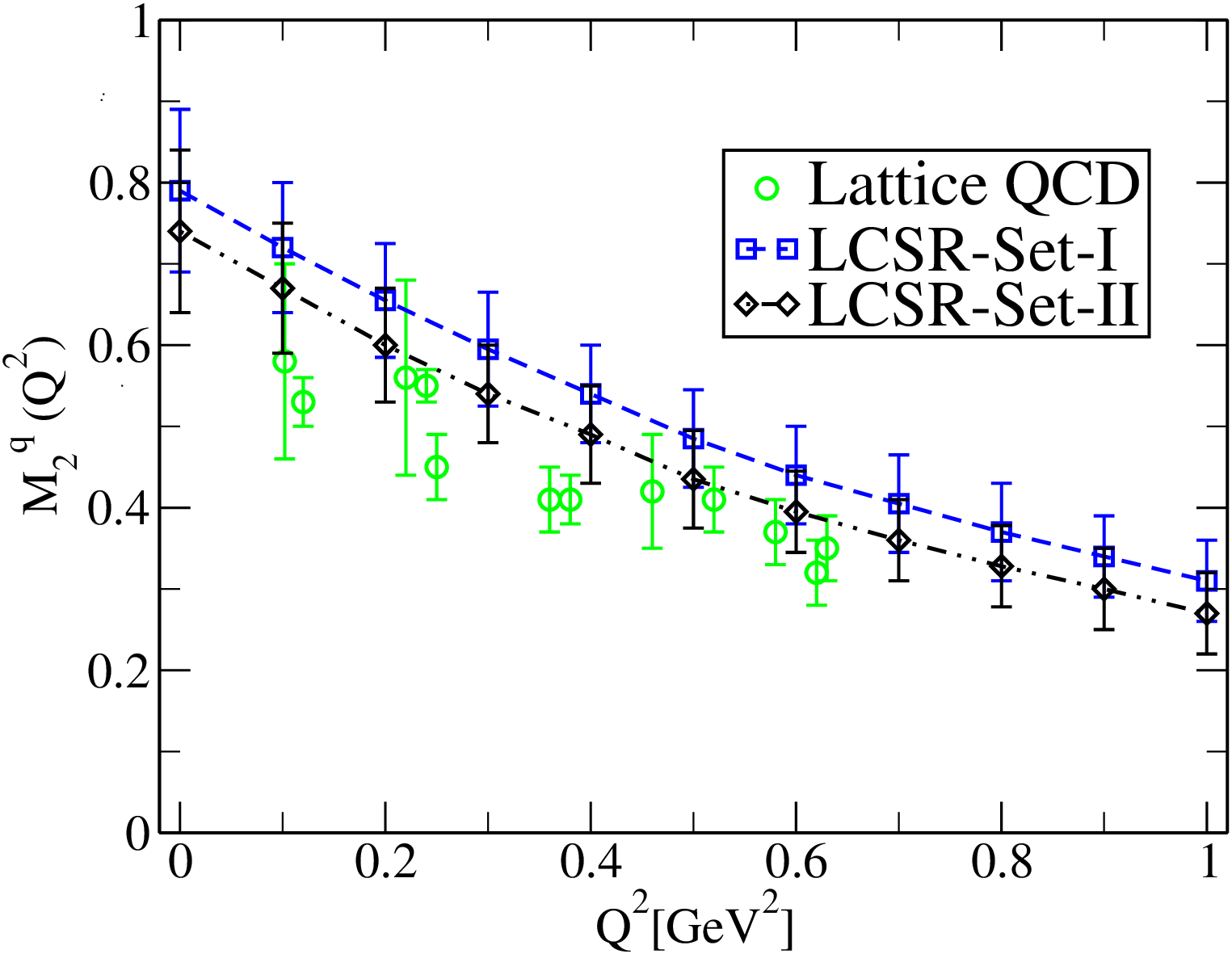}}\\
   \vspace{-0.1cm}
    \subfloat[]{\includegraphics[width=0.3\textwidth]{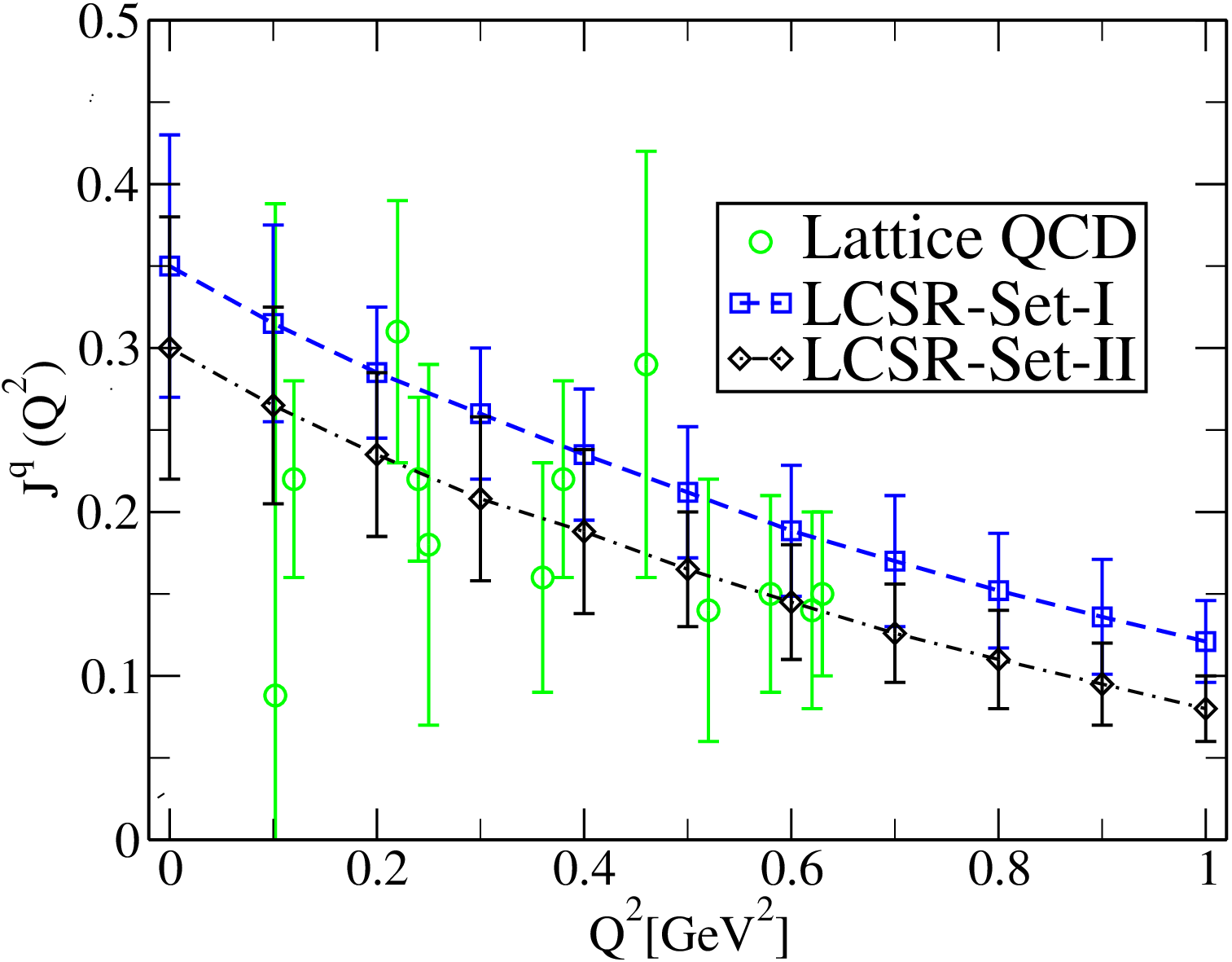}}\\
 \vspace{-0.1cm}
    \subfloat[]{\includegraphics[width=0.3\textwidth]{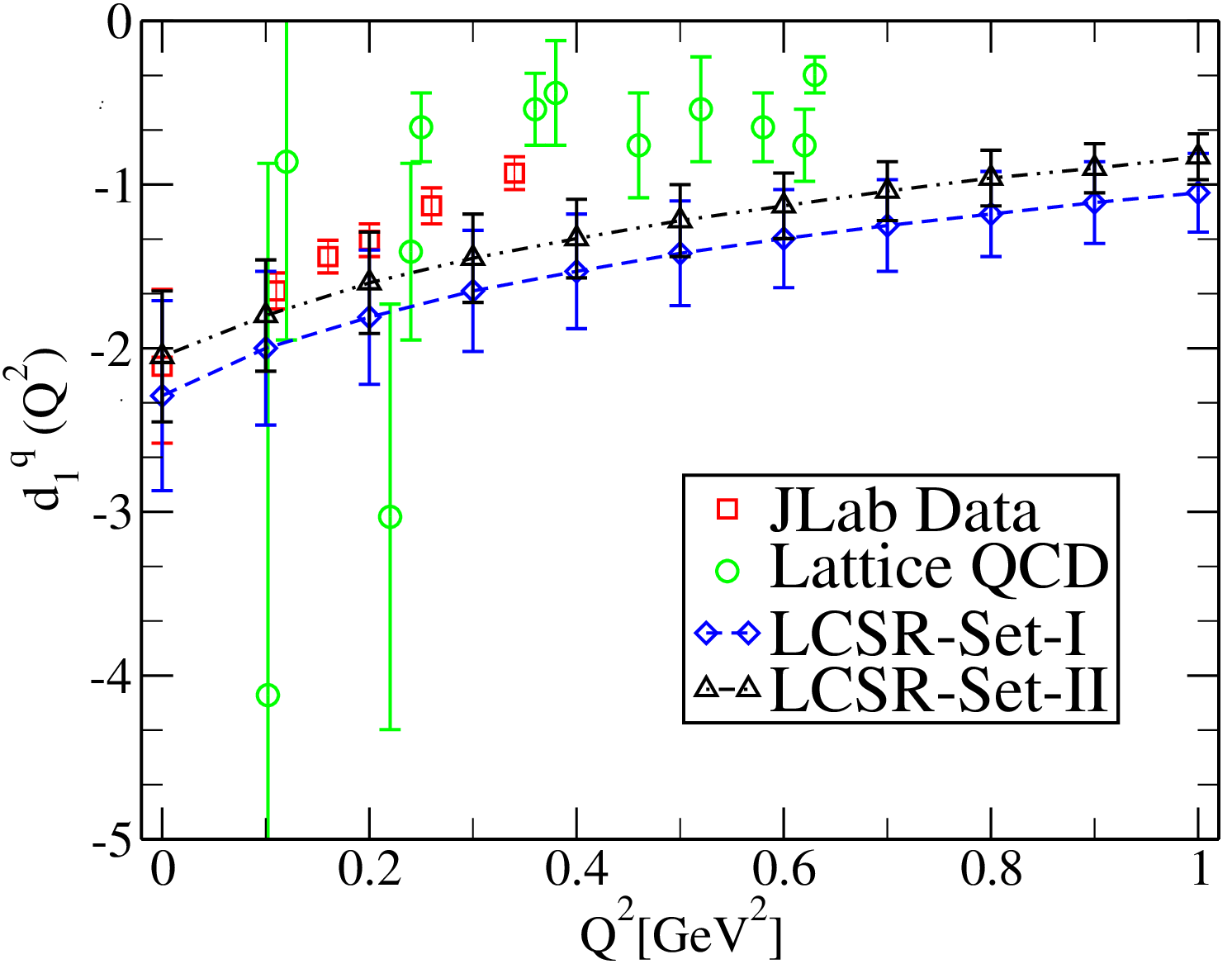}}
 \caption{The energy-momentum tensor form factors $M_2^q(Q^2)$, $J^q(Q^2)$ and $d_1^q(Q^2)$
 as functions of $Q^2$ at lower values of $ Q^2 $ and at the scale $\mu^2$ = 1.0 GeV$^2$. The Lattice data points are taken from the LHPC Collaboration~\cite{Hagler:2007xi} and JLab data  from ~\cite{Burkert:2018bqq}.}
 \label{comp}
  \end{figure} 

  \end{widetext}
Table \ref{compare} displays a comparison of our results at $Q^2=0$ with those of the  various theoretical models, Lattice QCD and existing experimental data for d$_1^q$(0) at re-normalization scale $\mu^2 = 1$ GeV$^2$.
For the M$_2 (0)$ form factor, almost all approaches give, more or less, similar predictions. 
For the J$^q(0)$ form factor, 
our estimations are in good agreements, within the errors, with the predictions of Refs.~\cite{Lorce:2018egm,Hagler:2003jd,mathur:1999uf,Gockeler:2003jfa, Bratt:2010jn, Hagler:2007xi,Brommel:2007sb,Negele:2004iu,Deka:2013zha,Dorati:2007bk}. 
Note that in the $\chi$QSM  and Skyrme models 2\,$J^q (0) = M_2^q$ = 1,  
since there are only quarks and antiquarks
to carry the nucleon's angular momentum and they must carry 100$\%$  of it.
%
Despite all of the results for d$_1^q$(0) from different sources have  the same sign, there are large discrepancies among the results. Our predictions for both of the DAs are in accord with the JLab data.
Our estimations are also consistent, within the errors, with the predictions of Refs.~\cite{Lorce:2018egm,Hagler:2007xi,Bratt:2010jn,Dorati:2007bk,Goeke:2007fp,Anikin:2019kwi,Pasquini:2014vua,Anikin:2017fwu}, but they differ   from the other predictions presented in the table.
The negative sign of the d$_1^q$(0) form factor indicates a profound connection with the  spontaneous 
chiral symmetry breaking in QCD (see also \cite{Polyakov:1999gs, Kivel:2000fg, Goeke:2001tz}) as well as an attractive relation with the criterion of stability of the nucleon~\cite{Goeke:2007fp}.
 The values obtained for the $\bar c^q (0)$ form factor using both of the DAs in the present study as well as the prediction of IP \cite{Polyakov:2018exb} are quite small, telling  us that the quark and gluon subsystems inside the nucleon interact weakly.
Our predictions for  $\bar c^q (0)$, however, differ with the predictions of IFF \cite{Lorce:2018egm}, substantially.   

At the end of this section, we would like to compare the behaviors of the $M_2^q(Q^2)$, $J^q(Q^2)$ and $d_1^q(Q^2)$ form factors at small values of $ (Q^2) $ with the Lattice predictions as well as JLab data for the $d_1^q(Q^2)$ gravitational form factor. To this end we plot Fig.~\ref{comp}.  The Lattice data points in this figure are taken from the LHPC Collaboration~\cite{Hagler:2007xi} and the JLab data form Ref.~\cite{Burkert:2018bqq}. For the form factor $M_2^q(Q^2)$, our predictions are consistent with most of the Lattice QCD data points considering the errorbars. In the case of $J^q(Q^2)$ and $d_1^q(Q^2)$ form factors the Lattice results suffer from large uncertainties at small values of $ Q^2 $. Our predictions, especially obtained via set-II DAs reproduce most of the JLab data at small 
 values of $ Q^2 $. This can be considered as a good assurance for the behaviors of the EMTFFs with respect to 
 $ Q^2 $ at all regions and in particular at $ Q^2=0 $.
 \begin{widetext}

\begin{table}[t]
	\addtolength{\tabcolsep}{2pt}
\begin{tabular}{l|c|c|c|ccc|c|cc}
				\hline\hline
				Approaches ~~~~~~&~~~~~~ M$_2^q$(0) ~~~~~~&~~~~~~ J$^q$(0) ~~~~~~&~~~~~~ d$_1^q$(0)~~~~~~&~~~~~~ $\bar c^q(Q^2)$\\
				\hline\hline
				Lattice QCD \cite{Hagler:2003jd}    & 0.74 $\pm$ 0.07 & 0.37 $\pm$ 0.03 & -&- \\
				Lattice QCD \cite{mathur:1999uf}    & 0.62 $\pm$ 0.14 & 0.31 $\pm$ 0.07 & -&-\\
				Lattice QCD \cite{Gockeler:2003jfa} & 0.72 $\pm$ 0.14 & 0.36 $\pm$ 0.08 & -1.09 $\pm$ 0.06&-\\
				Lattice QCD \cite{Bratt:2010jn}     & 0.59 $\pm$ 0.05 & 0.28 $\pm$ 0.02 & -1.76 $\pm$ 0.09 &-\\
				Lattice QCD \cite{Hagler:2007xi}    & 0.53 $\pm$ 0.02 & 0.28 $\pm$ 0.02 & -2.27 $\pm$ 0.30&-\\
				Lattice QCD \cite{Brommel:2007sb}   & 0.61 $\pm$ 0.15 & 0.24 $\pm$ 0.06 & - &- \\
				Lattice QCD \cite{Negele:2004iu}    & 0.74 $\pm$ 0.07 & 0.37 $\pm$ 0.03 & - &-\\
				Lattice QCD \cite{Deka:2013zha}     & 0.75 $\pm$ 0.07 & 0.38 $\pm$ 0.04 & - &-\\  
				$\chi$PT \cite{Dorati:2007bk}       & 0.57 $\pm$ 0.04 & 0.25 $\pm$ 0.06 & -1.93 $\pm$ 0.06 &- \\
				IFF \cite{Lorce:2018egm}            & 0.58 & 0.25 &-1.92 &-0.11\\
				Skyrme \cite{Cebulla:2007ei}        & 1 & 0.5 & -4.48 &-\\
				Skyrme \cite{Kim:2012ts}            & 1 & 0.5 & -3.54 &-\\				
				$\chi$QSM \cite{Goeke:2007fp}       & 1 & 0.5 & -2.35 &-\\
				$\chi$QSM \cite{Jung:2014jja}       & 1 & 0.5 & -5.03 & \\
				$\chi$QSM \cite{Wakamatsu:2007uc}   & - & -& -4.85 &- \\
				LCSR-LO \cite{Anikin:2019kwi}       & -&- & -2.63 $\pm$ 0.22 &- \\
				KM15 fit \cite{Anikin:2017fwu}      &-&-& -2.18 $\pm$ 0.21 &-\\
				DR \cite{Pasquini:2014vua}          & - & - & -1.70 &-\\
				JLab data \cite{Burkert:2018bqq}    &- &-& $ -2.11 \pm 0.46 $ &-\\
				IP \cite{Polyakov:2018exb}          & -&-&-&$ 1.1 \times 10^{-2}$\\
				This Work (Set-I)       & 0.79 $\pm$ 0.10 & 0.36 $\pm$ 0.10 & -2.29 $\pm$ 0.58 &$- (2.1\pm 0.8) \times 10^{-2}$ \\
				This Work (Set-II)      & 0.74 $\pm$ 0.12 & 0.30 $\pm$ 0.11 & -2.05 $\pm$ 0.40 &$-(2.5\pm 0.7)\times 10^{-2}$\\
      \hline\hline		
	\end{tabular}
\caption{The EMTFFs of nucleon at re-normalization scale $\mu^2 = 1.0$ GeV$^2$ compared with other predictions and JLab data. The Skyrme 
and $\chi$QSM models predictions were obtained considering both the quark and gluon parts of the EMT current and they are  re-normalization scale independent.}
	\label{compare}
\end{table} 

 \end{widetext}

\section{Mechanical Properties of Nucleon}\label{secIV}
Having calculated  the energy-momentum tensor form factors, it is straightforward to calculate  the pressure $p_0$ and the energy density ${\cal E}$ at the center of nucleon as well as estimate the hadron mechanical radius. The related formulas  for $p_0$ and ${\cal E}$  are given as~\cite{Polyakov:2018zvc}: 
\begin{eqnarray}
\label{Pres}
p_0  &=&-\frac{1}{24\,\pi^2\, m_N} \int^{\infty}_{0} dz \,z\,\sqrt{z}\, (d_1(z)-\bar c (z) ),\\
%
{\cal E}&=&\frac{m_N}{4\,\pi^2} \int^{\infty}_{0} dz \,\sqrt{z}\,\Big[ M_2(z) + \frac{z}{4m^2_N} [M_2(z) -2J(z)+ d_1(z)+ \bar c (z)]\Big],
\end{eqnarray}
where $z=Q^2$. Our results  on the mechanical quantities $ p_0 $  and $ {\cal E} $ of the nucleon compared to other existing theoretical predictions are shown in table \ref{mech_table1}. 
One can see from Table \ref{mech_table1} that our predictions on $ p_0 $, within the errors, are very close to that of \cite{Jung:2014jja}, however, they differ with other predictions, considerably. Our results on $ {\cal E} $ using both sets of DAs are close to the predictions of \cite{Kim:2012ts} and  \cite{Goeke:2007fp}, but demonstrate considerable deviations from other presented predictions. We should remind that $ \bar c ^q(z)+ \bar c ^g(z) =0$, in the case one considers both the gluonic and quark parts of the energy-momentum tensor, implying that the energy-momentum tensor current is conserved. However, as we previously mentioned we obtained  very small values for the form factor  $ \bar c ^q(z)$ at different points, referring to a very small violation of the current conservation. These small values for  $ \bar c ^q(z)$ do not affect the mechanical properties. We shall also note that the quantity $ M_2(z) -2J(z) $ has very small impact on the ${\cal E}  $, a result that also is found in \cite{Polyakov:2018zvc}. Ignoring $\bar c ^q(z)  $, for the mechanical mean squre radius, one obtains ~\cite{Polyakov:2018zvc}
\begin{eqnarray}
\langle r^2_{\text{mech}}\rangle = 6\, d_1(0) \Big[ \int^{\infty}_{0} dz\, d_1(z)\Big]^{-1}.
\end{eqnarray}
The numerical results for $ \langle r^2_{\text{mech}}\rangle  $, using two sets of DAs, compared to the only existing prediction from Ref.  \cite{Anikin:2019kwi} are shown in table \ref{mech_table1}, as well.  As it is seen, our predictions on $ \langle r^2_{\text{mech}}\rangle $ using both sets are in good consistencies with the  prediction of Ref. \cite{Anikin:2019kwi} within the presented uncertainties. The presented results and their comparison with probable future experimental data   can be very useful in understanding the structure of the nucleon.

 \begin{widetext}

\begin{table}[htp]
\hspace*{-0.5cm}
\begin{tabular}{ l|c|c|c|c|c|c|cc}
 \hline\hline
Mechanical properties  &Results of set-I & Results of set-II & \cite{Cebulla:2007ei}& \cite{Kim:2012ts}&  \cite{Goeke:2007fp}&  \cite{Jung:2014jja}& \cite{Anikin:2019kwi}\\
\hline\hline
      $p_0$  (GeV/fm$^3$)                                     &$0.67 \pm 0.09$ & $0.62 \pm 0.08 $   &0.47 & 0.26 & 0.23 & 0.58 & 0.86 \\
      ${\cal E}$ (GeV/fm$^3$)                                  &$1.76 \pm 0.18$ & $1.74 \pm 0.14 $   &2.28 & 1.45 & 1.70 & 3.56 & 0.94  \\
      $\langle {r^2_{\text{mech}}\rangle}$ (fm$^{2}$)          &$0.54   \pm 0.06$ & $0.52 \pm 0.05 $ &    -&-     & - &-& 0.54  \\
\hline \hline
\end{tabular}
\caption{The values of mechanical quantities of nucleon.}
	\label{mech_table1}
\end{table}

 \end{widetext}

\section{Summary and Concluding Remarks}\label{secV}

The energy-momentum tensor or gravitational form factors of nucleon are basic quantities that carry valuable information on different aspects of the nucleon's structure. These are used to calculate  the pressure and energy distributions inside the nucleon as well as quantities related to its geometric shape. 
The EMTFFs are also sources of information on the fractions of the momenta carried by the quarks and gluons as ingredients of the nucleon. They help us  know  how the total angular momenta of quarks and gluons form the nucleon's spin. They also provide knowledge on the distribution and stabilization of the strong force inside the nucleon. We  extracted  the EMTFFs of the nucleon by applying the light-cone QCD sum rule formalism and using two different sets of the parameters inside the nucleon's distribution amplitudes. In the calculations, we used the most general interpolating current of the nucleon in terms of its constituent quark fields. We observed that the results don't depend on the choice of the DAs, considerably and the two sets give  close results to each other. We found that the EMTFFs of nucleon are best described by a multipole fit function, helped us to extrapolate the results to the regions that the LCSRs results are not reliable and applicable. 

We extracted the numerical values of the EMTFFs at $ Q^2=0 $ and compared the results with the existing theoretical predictions as well as the results of the Lattice QCD and JLab data. We observed a consistency among the theoretical predictions, within the uncertainties, for the values of the form factors M$_2^q (0)$ and J$^q(0)$. Our results on these form factors are nicely consistent with the Lattice QCD and chiral perturbation theory predictions. However, there are large discrepancies among the theoretical predictions on the gravitational form factor d$_1^q(0)$.  Nevertheless, our prediction is  in accord with the JLab data as well as
with the predictions of the Lattice QCD, chiral perturbation theory and KM15-fit. We obtained a very small value for the $\bar c^q (0)$ form factor referring to a good conservation of the quark part of the  EMT current.  

We discussed the behavior of the EMTFFs with respect to $ Q^2 $ and observed that all form factors approach to zero at large values of  $ Q^2 $. Making use of the multipole fit function, we glanced the behavior of the FFs at small values of $ Q^2 $, where we have some experimental data on d$_1^q(Q^2)$ provided by JLab. We saw that
the fit function considered in the present study well defines most of the JLab data in the interval $ Q^2\in[0,0.4]~GeV^2 $. The Lattice QCD results for d$_1^q(Q^2)$ and J$^q(Q^2)$ suffer  from large uncertainties in this region. The behaviors of the form factor M$_2^q (Q^2)$ obtained using two sets of DAs in the present study are  consistent with the Lattice QCD predictions that contain small uncertainties at small values of $ Q^2 $.

Making use of the fit functions of the form factors, we calculated   the pressure and  energy density distributions  at the center of nucleon as well as the mechanical radius of the nucleon and compared with the existing theoretical predictions.   Our predictions on $ p_0 $, using two sets of DAs, are very close to that of $\chi$QSM \cite{Jung:2014jja} within the errors, however, they differ with other predictions presented in table \ref{mech_table1}, considerably. Our predictions on $ {\cal E} $ at the center of the nucleon are close to the predictions of Skyrme model\cite{Kim:2012ts} and $\chi$QSM \cite{Goeke:2007fp}, but demonstrate considerable deviations from other presented predictions. The predictions of the present study on $ \langle r^2_{\text{mech}}\rangle $  are in good consistencies with the only existing prediction provided by LCSR-LO approach \cite{Anikin:2019kwi} within the presented uncertainties.

The presented results in this study together with the predictions of Lattice QCD and other theoretical predictions on the nucleon's EMTFFs may help experimental groups to measure the values of these form factors at a wide range of $ Q^2 $. The good consistency between our predictions and the existing JLab data on d$_1^q(Q^2)$ in the interval $ Q^2\in[0,0.4]$~GeV$^2 $, strengthens this hope. Any experimental data on the energy momentum tensor as well as the electromagnetic, axial and other form factors of the nucleon and their comparison with the theoretical predictions can help us gain valuable knowledge on the internal structures and geometric shapes of the nucleons as building blocks of the visible matter. Such investigations may also help us  answer many fundamental questions by means of the quark-gluon structures of the nucleons.

\section{Acknowledgements}
U. \"{O}. thanks the scientific and technological research council of Turkey (TUBITAK) for the support provided under 2218-National Postdoctoral Research Scholarship Program.

\begin{widetext}
 \appendix
\section{ Alternative definition of the EMT current's matrix element}
By exploring the Gordon equality $2M_N\bar u^\prime\gamma^\alpha u=$ $\bar u^\prime(i\sigma^{\alpha\beta}\Delta_\beta+2P^\alpha)u$ an alternative decomposition of  Eq. (\ref{mat}) is obtained:
 \begin{eqnarray}\label{altmat}
\langle N(p^\prime,s')|T_{\mu\nu}|N(p,s)\rangle&=&
 \bar{u}(p^\prime,s')\Big[A(Q^2)\frac{ \gamma_\mu P_\nu  + \gamma_\nu P_\mu}{2}
 +B(Q^2)\frac{i(\tilde P_\mu \sigma_{\nu\rho}+\tilde P_\nu \sigma_{\mu\rho})\Delta^\rho}{4m_N}
 +C(Q^2) \frac{\Delta_\mu \Delta_\nu- g_{\mu\nu} \Delta^2}{m_N}
   \nonumber\\
  && +~ \bar C (Q^2)  m_N g_{\mu \nu} \Big] u(p,s), 
\end{eqnarray}
where
\begin{align}
A(Q^2)  &= M_2(Q^2),\nonumber\\
A(Q^2)+B(Q^2))&= 2\,J(Q^2),\nonumber\\
C(Q^2)   &= \frac15\,d_1(Q^2)\nonumber.
\end{align}

\section{ Explicit forms of the F functions  for the \texorpdfstring{$M_2^q(Q^2)$}{} form factor}
\begin{align*}
 F_1(x_2) =&-2 \int_0^{1-x_2} dx_1 (1+t)[P_1+S_1](x_1,x_2,1-x_1-x_2),\nonumber\\
 F_2(x_3) = & 2  \int_0^{1-x_3} dx_1[2 (1-t)(V_3-A_3)- (1+t)(P_1+S_1+2T_1-4 T_7)](x_1,1-x_1-x_3,x_3),\nonumber\\
 F_3(x_2) = & 2 \int_0^{1-x_2} dx_1[(1-t)(A_1+V_1)-(1+t)(P_1+S_1)](x_1,x_2,1-x_1-x_2),\nonumber\\
 F_4(x_3) = & 2 \int_0^{1-x_3} dx_1[(1-t)(A_1-2A_3+V_1+2V_3)-(1+t)(P_1+S_1-2 T_1+4 T_7)](x_1,1-x_1-x_3,x_3),\nonumber\\
 F_5(x_2) = &-2 \int_0^{1-x_2} dx_1 [5(1-t)( A_1^{M}+ V_1^M)+(1+t)T_1^M](x_1,x_2,1-x_1-x_2),\nonumber\\
 F_6(x_3) = &-2 \int_0^{1-x_3} dx_1 [(1-t)( A_1^M+ V_1^M)-2 (1+t)T_1^M](x_1,1-x_1-x_3,x_3),\nonumber\\
 F_7(\alpha) = & \int_\alpha^1 dx_2 \int_0^{1-x_2} dx_1 [2(1-t)(A_1-A_2+A_3+V_1-V_2-V_3)-(1+t)(T_1+3 T_2-4 T_3+2 T_7)]\nonumber\\
 &(x_1,x_2,1-x_1-x_2),\nonumber\\
 F_8(\alpha) = &  \int_\alpha^1 dx_3 \int_0^{1-x_3} dx_1 [(1+t)(-5T_1-T_2+6 T_3+2 T_7)](x_1,1-x_1-x_3,x_3),\nonumber\\
  F_9(\alpha) = & \int_\alpha^1 dx_2 \int_0^{1-x_2} dx_1 [(1-t)(-11T_1+T_2+5T_3+12T_7)](x_1,x_2,1-x_1-x_2),\nonumber\\
 F_{10}(\alpha) = &  \int_\alpha^1 dx_3 \int_0^{1-x_3} dx_1 [(1+t)(-11T_1-3T_2+14T_3+T_7)](x_1,1-x_1-x_3,x_3),\nonumber
 \end{align*}
\begin{align*}
 F_{11}(\alpha) = & \int_\alpha^1 dx_2 \int_0^{1-x_2} dx_1 [4(1-t)(-A_3+A_4+V_3-V_4)+(1+t)(-4P_1+4P_2-4S_1+4S_2+4T_2-T_3+4T_5\nonumber\\
 &+T_7)](x_1,x_2,1-x_1-x_2),\nonumber\\
 F_{12}(\alpha) = &  \int_\alpha^1 dx_3 \int_0^{1-x_3} dx_1 [4(1-t)(-A_1+A_2-A_3+A_4-V_1+V_2+V_3-V_4)+(1+t)(-4P_1+4P_2-4S_1+4S_2\nonumber\\
&-T_1+T_2-4T_3+4T_5+10T_7)](x_1,1-x_1-x_3,x_3),\nonumber\\
 F_{13}(\alpha) = & \int_\alpha^1 dx_2 \int_0^{1-x_2} dx_1 [4(1-t)(-A_3-A_4+V_3-V_4)+(1+t)(4P_2-4S_1+4S_2+4T_2-T_3+4T_5+T_7)]\nonumber\\
&(x_1,x_2,1-x_1-x_2),\nonumber\\
F_{14}(\alpha) = &  \int_\alpha^1 dx_3 \int_0^{1-x_3} dx_1 [4(1-t)(-A_1-A_2-A_3+A_4-V_1+V_2+V_3-V_4)+(1+t)(-4P_1+4P_2-4S_1+4S_2\nonumber\\
&-T_1+T_2-4T_3+4T_5+10T_7)](x_1,1-x_1-x_3,x_3),\nonumber\\
F_{15}(\beta) = &2 \int_0^\beta d\alpha\int_\alpha^1 dx_2 \int_0^{1-x_2} dx_1[(1+t)(-7 T_1+2T_2+5T_3+5T_4+2T_5-7T_6-9T_7+9T_8)](x_1,x_2,1-x_1-x_2),\nonumber\\
F_{16}(\beta) = & \int_0^\beta d\alpha\int_\alpha^1 dx_3 \int_0^{1-x_3} dx_1 [(1+t)(-T_1+2T_2+6T_3+6T_4+2T_5-T_6+10T_7+10T_8)](x_1,1-x_1-x_3,x_3),\nonumber\\
F_{17}(\beta) = &2 \int_0^\beta d\alpha\int_\alpha^1 dx_2 \int_0^{1-x_2} dx_1[(1+t)(-7 T_1+T_2+3T_3+3T_4+T_5-7T_6-8T_7+8T_8)](x_1,x_2,1-x_1-x_2),\nonumber\\
F_{18}(\beta) = & \int_0^\beta d\alpha\int_\alpha^1 dx_3 \int_0^{1-x_3} dx_1 [(1+t)(-T_1-4T_2+6T_3+6T_4-4T_5-T_6+4T_7+4T_8)](x_1,1-x_1-x_3,x_3),\nonumber
\end{align*}
\end{widetext}

\bibliography{refs}

\end{document}